\documentclass[prb,twocolumn]{revtex4-1}

\usepackage{bm}
\usepackage{xr}
\usepackage{graphicx}
\usepackage{amsmath}
\usepackage[margin=1.63cm]{geometry}
\usepackage{amssymb}

\usepackage{caption}
\newcommand*{\cuacac}{$[$Cu(acac)$_2]$}
\newcommand*{\jp}{J_{\parallel}}
\newcommand*{\jone}{J_{\perp 1}}
\newcommand*{\jtwo}{J_{\perp 2}}

\begin{document}

\title[]
{Towards mechanomagnetics in elastic crystals: insights from [Cu(acac)$_2$]}

\author{E. P. Kenny}
\affiliation{School of Mathematics and Physics, The University of Queensland, Brisbane, Queensland, Australia}
\email{elisekenny@gmail.com}
\author{A. C. Jacko}
\affiliation{School of Mathematics and Physics, The University of Queensland, Brisbane, Queensland, Australia}
\author{B. J. Powell}
\affiliation{School of Mathematics and Physics, The University of Queensland, Brisbane, Queensland, Australia}

\begin{abstract}
We predict that the magnetic properties of \cuacac, an elastically flexible crystal, change dramatically when the crystal is bent. We find that unbent \cuacac\ is an almost perfect Tomonaga-Luttinger liquid. Broken-symmetry density functional calculations reveal that the magnetic exchange interactions along the chains is an order of magnitude larger than the interchain exchange. The geometrically frustrated interchain interactions cannot magnetically order the material at any experimentally accessible temperature. The ordering temperature ($T_N$), calculated from the chain random phase approximation, increases by approximately 24 orders of magnitude when the material is bent. We demonstrate that geometric frustration both suppresses $T_N$ and enhances the sensitivity of $T_N$ to bending. In \cuacac, $T_N$ is extremely sensitive to bending, but remains too low for practical applications, even when bent.   
Partially frustrated materials could achieve the balance of high $T_N$ and good sensitivity to bending required for practical applications of mechanomagnetic elastic crystals.
\end{abstract}

\maketitle

\section*{Introduction}
Crystal adaptronics is a new and exciting field, bolstered by the recent discovery of elastically flexible molecular crystals.\cite{Chen2014, Ghosh2015jacs, Ghosh2015angew, Hayashi2016, Worthy2017, Hayashi2018, Ahmed2018} These crystals can be bent without irreversibly changing their structure. The mechanism by which the molecules can elastically slip past each other is beginning to be understood.\cite{Reddy2006, MishraReview, Saha2018, Brock2018, Zakharov2019} However, there are limited examples of the modification of functional properties; the most successful so far being mechanochromism.\cite{Hayashi2016, Hayashi2017, Hayashi2018angew} 

In this paper, we explore the possible changes in  magnetic properties induced by bending \cuacac\ (acac$=$acetylacetonate), a recently discovered elastically flexible crystal. \cite{Brock2018, Worthy2017} We discuss how the geometry of the crystal leads to these changes in the hope of motivating a search for elastic crystals with similar geometry, but larger exchange interactions.

\cuacac\ is an extremely well known material. It is a commercially available reactant used in numerous organic and organometallic syntheses and is often made in undergraduate chemistry laboratories. Worthy \textit{et al.}\cite{Worthy2017} published atomically resolved structural information across bent samples, providing the opportunity to use first-principles calculations to model how its magnetic properties change as the material is bent.

We find that, apart from being elastic, \cuacac\ has exotic quantum magnetic properties -- it is an almost perfect quasi-one-dimensional magnet. The frustrated geometry of the crystal lattice enhances this low dimensionality and also leads to extreme sensitivity of the magnetic properties to bending. \cuacac 's partnership of elasticity and geometrical frustration lead to it being an excellent prototype for applications for elastic crystals. We predict that the change in geometry of \cuacac, brought on by bending, will lead to its magnetic ordering temperature changing by approximately 24 orders of magnitude. This demonstrates the possibility of using elastic flexible crystals to passively sense small deformations or flexures with extremely high precision.

Passive flex sensors often operate with a change in electrical resistivity. They are useful for measuring physical activity or joint movement in the human body, for facilitating human-computer interactions, for monitoring machines, and for measurement devices (for example measuring the curvature of a small surface).\cite{Saggio2016} A material with dramatic magnetic changes caused by bending, such as \cuacac, could also be used for these purposes down to the micrometer scales. The magnetic ordering temperature, which can be detected via the concomitant divergence in the magnetic susceptibility, can change by many orders of magnitude; such devices could have sensitivities far exceeding those of resistive devices.

The behavior of flexible quantum magnets, a new field opened by the discovery of elastic crystals, allows one to examine many new questions of fundamental importance.
Low dimensional magnetic crystals display fascinating quantum phenomena. Particularly, one-dimensional materials exhibit  fractionalized excitations and strong quantum fluctuations that prohibit long range magnetic order.\cite{Balents2010, BalentsReview} Spin-1/2 one-dimensional Heisenberg chains are described by Tomonaga-Luttinger liquid (TLL) theory. \cite{Tomonaga1950, Luttinger1963,Haldane1981} An important prediction of TLL theory is that there will be a continuum of low-energy excitations, which are indeed observed in neutron scattering experiments. \cite{Lake2005} 
Quasi-one-dimensional crystals contain weak interchain interactions, which become significant at low temperatures and lead to N\'eel ordering below a certain temperature, $T_N$. These materials can be understood as weakly coupled chains. However, at low enough temperatures, interchain interactions eventually cause long-range magnetic order.

Copper II molecular crystals are well known for their exotic magnetic properties. \cite{Landee2013} One of the best examples of a quasi-one-dimensional molecular crystal is copper pyrazine dinitrate, [Cu(pz) (NO$_3$)$_2$] (pz$=$pyrazine), which orders magnetically at 0.107 K \cite{Lancaster2006} and was recently shown to exhibit 1D quantum criticality. \cite{Breunig2017} Its magnetic low dimensionality has been confirmed with density functional theory calculations, which give an interchain coupling of $J_\perp=0.0044\jp$, where $\jp$ is the intrachain coupling.\cite{Somoza2010} The extent to which a material is 1D can be quantified with $T_N/\jp$. Table \ref{tab:other_tns} shows some of the lowest values found to date.

\begin{table}
\caption{The ordering temperatures in units of $\jp$ of various quasi-one-dimensional molecular crystals found in the literature along with the value predicted for \cuacac\ in this paper. A lower value of $T_N/\jp$ indicates a material closer to the 1D limit.}\label{tab:other_tns}
\begin{center}
\begin{tabular}{c|cc}
Material & $T_N/\jp$ & Ref. \\ 
\hline 
Cu$_3$(CO$_3$)$_2$(OH)$_2$ & 0.19 & \citenum{Rule2011} \\
Cs$_2$CuCl$_4$ & 0.15 & \citenum{Coldea2001} \\ 
KCuF$_3$ & 0.10 & \citenum{Lake2005, Satija1980, Hutchings1969} \\
$[$Cu(pz)(pyO)$_2$(H$_2$O)$_2]$ (PF$_6$)$_2$ & 0.03 & \citenum{Goddard2012} \\
$[$Cu(pz)(NO$_3$)$_2]$ & 0.01 & \citenum{Lancaster2006} \\
\cuacac & $10^{-33}$ & This work
\end{tabular}
\end{center}
\end{table}

\begin{figure}
\centering
\includegraphics[width=7cm]{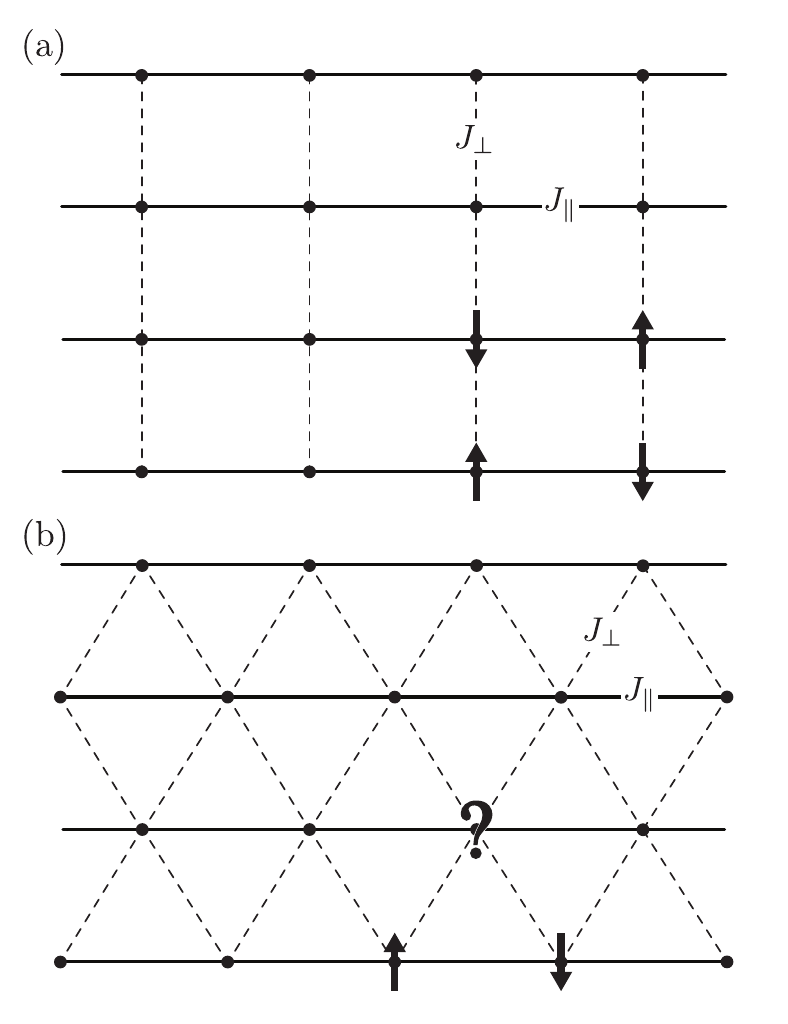}
\caption{Two examples of coupled chain geometries; (a)  perpendicular interchain couplings and (b)  frustrated triangular couplings. Most of the materials in Table \ref{tab:other_tns} have some combination  of both of these types of interactions; however, \cuacac, only has frustrated interactions (b). This is why $T_N/\jp$ is much lower in \cuacac. In these quasi-1D materials, $\jp$ (within the spin chains) is strongly antiferromagnetic, favoring short-range antiferromagnetic correlation, as shown. The triangular geometry in (b) frustrates the interchain couplings, $J_{\perp}$, as indicated by `?' (regardless of whether the interchain couplings are ferromagnetic or antiferromagnetic). Whereas the square geometry in (a) is unfrustrated. }
\label{fig:frustration}
\end{figure}

Below, we demonstrate that unbent \cuacac\ is an almost perfect TLL that does not order magnetically at any experimentally accessible temperature.
We establish this through a combination of first principles electronic structure calculations and quantum many-body theory, revealing that the presence of geometrical frustration in the lattice (see Fig. \ref{fig:frustration}) causes two major effects: (i) \cuacac's extreme magnetic one-dimensionality and (ii) the significant change the N\'eel temperature, $T_N$, when the material is bent. 

We parametrize a Heisenberg Hamiltonian via broken-symmetry density functional theory (BS-DFT), \cite{Noodleman1981, Mouesca} which reveals three significant exchange couplings between neighboring molecules, $\jp$, $\jone$, and $\jtwo$ (shown in Fig. \ref{fig:J_labels}). The magnitude of the exchange coupling along the crystallographic $b$-axis ($\jp$) is much larger than the couplings in the other directions, indicating that \cuacac\ can be modelled as weakly coupled Heisenberg spin-1/2 chains. The interchain couplings, $\jone$ and $\jtwo$, are both geometrically frustrated (see Figs. \ref{fig:frustration} and \ref{fig:J_labels}), maintaining \cuacac\ in the 1D limit.

We use the chain random phase approximation (CRPA) \cite{Bocquet2001} to predict the N\'eel temperature, magnetic susceptibility and dynamical structure factor of the unbent crystal.  The measured susceptibility is in good agreement with our calculations. When the crystal is bent, the ratio of intra to interchain couplings changes significantly. This leads to a change in N\'eel temperature of 24 orders of magnitude, demonstrating the dramatic potential of mechanomagnetics.

\section*{Computational Details and Theoretical Methods}

\begin{figure}
\centering
\includegraphics[width=8cm]{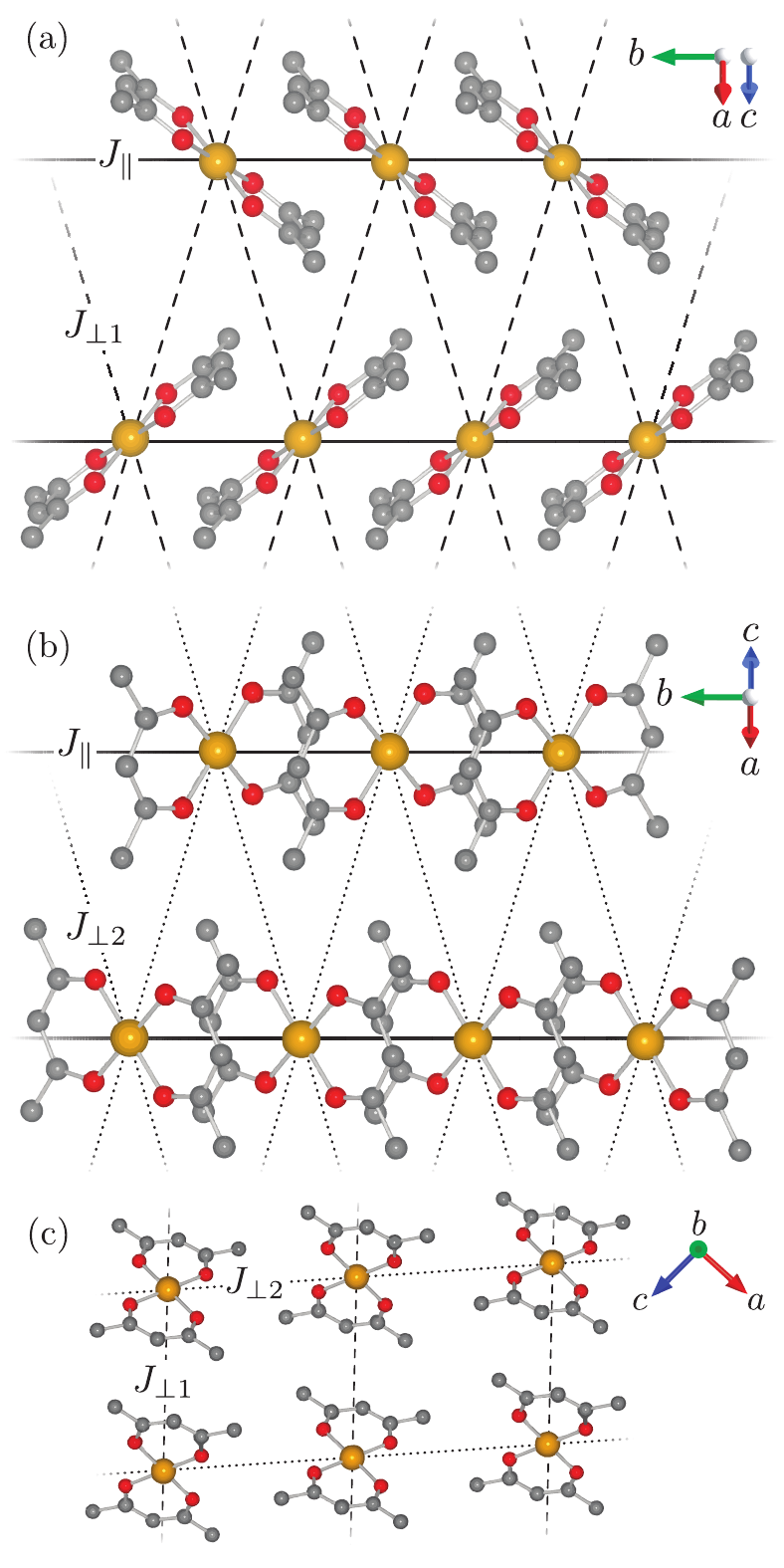}
\caption{The nearest neighbor exchange pathways in \cuacac. The crystallographic axes ($a$, $b$, and $c$) are shown. The lattice is geometrically frustrated. We find that $\jp$ is strongly antiferromagnetic, favoring short-range antiferromagnetic correlation. Regardless of the signs of $\jone$ and $\jtwo$ (i.e. whether they are antiferromagnetic or ferromagnetic), the triangular geometry frustrates these couplings.}
\label{fig:J_labels}
\end{figure}

We use the unbent and bent \cuacac\ crystal structures from Worthy \textit{et al.}\cite{Worthy2017} Three nearest neighbor exchange pathways are shown in Fig. \ref{fig:J_labels}. In terms of the crystallographic axes we label $\jp$ to be along $b$. The four nearest neighbour interactions in the $\pm\left(b/2\pm\left(a+c\right)\right)$ directions are equal (by symmetry) and we label them $\jone$. Similarly, we label the four  nearest neighbour interactions along $\pm\left(b/2\pm\left(a-c\right)\right)$ as $\jtwo$. 

When the crystal is bent, the lattice parameters change approximately linearly as a function of position across the bend.\cite{Worthy2017} On the inside of the bend, the $b$-axis is compressed while the $a$ and $c$ axes are stretched. Conversely, on the outside of the bend, the $b$-axis stretches while $a$ and $c$ are compressed. The $\beta$ angle increases approximately linearly from the outside to the inside of the bend. However, the individually measured atomic coordinates are not as precise as those from bulk crystals due to the small effective sample size. We therefore created a linearized set of lattice parameters using crystallographic data for two bends with different radii of curvature, $r_c=1.2\,$mm and $r_c=3.2\,$mm. We then used these parameters to produce a new smooth set of model structures assuming constant intra-molecular bond lengths. Details of this process and a plot of the lattice parameters across each bend, including our linear fits, are given in the Supplementary Information.

We parametrize a Heisenberg model,
\begin{equation}
\mathcal{H}_\mathrm{Heisenberg}=\sum_{ ij} J_{ij}\ \bm{S}_i\cdot\bm{S}_j,
\label{eq:HHam}
\end{equation} 
where $\bm{S}_i$ is the spin operator on the $i$th molecule  and $J_{ij}$ are the exchange coupling constants. The sign of $J$ indicates an antiferromagnetic ($J\!>\!0$) or ferromagnetic ($J\!<\!0$) interaction. 

We calculate the exchange couplings, $J_{ij}$, within \cuacac\ using broken-symmetry density functional theory (BS-DFT),\cite{Noodleman1981, Mouesca} along with the Yamaguchi spin decontamination procedure.\cite{Yamaguchi1988} In this approach, 
\begin{equation}
J_{ij}=2\frac{E^\mathrm{BS}_{ij}-E^\mathrm{T}_{ij}}{\langle S^2\rangle^\mathrm{BS}_{ij}-\langle S^2\rangle^\mathrm{T}_{ij}},
\label{eq:BS}
\end{equation}
where $E^\mathrm{T}_{ij}$ is the triplet energy of the isoloated dimer containing molecules $i$ and $j$, and $E^\mathrm{BS}_{ij}$ is the energy of the broken-symmetry state on that same dimer. $\langle S^2\rangle^\mathrm{BS}_{ij}$ and $\langle S^2\rangle^\mathrm{T}_{ij}$ are the corresponding expectation values of the spin operator, $S^2$. Calculations were perfomed in Gaussian09 \cite{g09} with the uB3LYP functional \cite{B3LYPa,B3LYPb} and using the LANL2DZ\cite{Dunning1977, Hay1985a, Wadt1985, Hay1985b} (for Cu) and 6-31+G*\cite{Pople1, Pople2, Pople3, Pople4} basis sets with an SCF convergence criterion of $10^{-10}$\,a.u. Benchmarking of $J_{ij}$ using different basis sets and functionals is discussed in the Supplementary Information.

The dynamical magnetic susceptibility for a single Heisenberg chain can be calculated from a combination of the Bethe ansatz and quantum field theory techniques. \cite{Bethe1931, Yang1966, Schulz1983, Eggert1994, Schulz1986, Affleck1998, Barzykin2000, Lukyanov1998, Luther1974, TsvelikBook} Within the chain random phase approximation (CRPA), the full three-dimensional dynamical susceptibility is a function of the interchain coupling, $J_\perp$ (see Eq. S.3).\cite{Scalapino1975, Schulz1996, Essler1997, Bocquet2001} The CRPA susceptibility is valid above the N\'eel temperature, $T_N$. Generically, one expects an RPA treatment to overestimate $T_N$. However, the geometrical frustration in \cuacac\ enhances the range of validity of this approximation; the CRPA has been compared with numerical methods and found to be accurate for $|J_\perp|<0.1\,\jp$ on a geometrically unfrustrated lattice \cite{Yasuda2005} and $|J_\perp|<0.7\,\jp$ for a frustrated lattice.\cite{Starykh2010} One can determine $T_N$ by considering the condition for a zero frequency pole in the CRPA expression for the dynamical susceptibility. Details of this calculation are given in the Supplementary Information.

We use the CRPA to predict a number of experimentally measurable properties of \cuacac. We fit the CRPA, using the Bonner-Fisher chain susceptibility\cite{Bonner1964, Estes1977, Ami1995} to the experimental bulk susceptibility above 2\,K.\cite{Moreno2013} We then predict the low-temperature CRPA susceptibility ($T_N<T<1.5$\,K) with the temperature dependent bulk susceptibility of a single antiferromagnetic Heisenberg chain, calculated numerically by Eggert \textit{et al}.\cite{Eggert1994,EggertDat}  The bulk magnetic susceptibility will diverge, undergoing a second order phase transition, at $T_N$. The dynamical structure factor (measured in inelastic neutron-scattering experiments) can also be calculated with the CRPA susceptibility (see S.19). Details of the experimental predictions are also given in the Supplementary Information.

\section*{Results and Discussion}

\textit{Unbent Crystal.} The three distinct BS-DFT nearest neighbor exchange interactions in the unbent crystal, along with their crystallographic directions, are reported in Table \ref{tab:DFT_results}. All longer range interactions that we calculated are smaller than the accuracy limit of our DFT results. 

\begin{table}
\caption{Heisenberg exchange ($J_{ij}$) parameters for the unbent structure of \cuacac\, determined with BS-DFT. $\jp$ and $\jone$ are antiferromagnetic and $\jtwo$ is ferromagnetic. The distances between Cu atoms for each dimer are also reported. Axes are shown in Fig. 
\ref{fig:J_labels}.}\label{tab:DFT_results}
\begin{center}
\begin{tabular}{c|ccc}
  & Direction & Cu$\leftrightarrow$Cu\,(\AA)  & $J_{ij}/k_B$\,(K)\\ 
\hline 
$\jp$ & $\pm\bm{b}$ & 4.643 & 0.75\\ 
$\jone$ & $\pm\bm{b}/2\pm\left(\bm{a}+\bm{c}\right)$&  7.818 & 0.04  \\ 
$\jtwo$ & $\pm\bm{b}/2\pm\left(\bm{a}-\bm{c}\right)$&  8.133 & -0.10   
\end{tabular}
\end{center}
\end{table}

The exchange coupling ratios, $J_{\perp 1}/J_{\parallel} = 0.06,\ J_{\perp 2}/J_{\parallel} = -0.13,$ indicate a low dimensionality in the magnetic degrees of freedom in \cuacac. 
In the limit $\jone=\jtwo=0$, one has independent Heisenberg chains which are Tomonaga-Luttinger liquids (TLLs) at low temperatures. \cite{Haldane1980} However, when there are interactions between chains (i.e. $\jone,\jtwo\neq0$), the TLL will undergo a phase transition into a N\'eel ordered state below a critical temperature, $T_N$. 

Using the CRPA susceptibility (details given in the Supplementary Information), we find that the N\'eel temperature of \cuacac\ is given by 
\begin{equation}
T_N \approx \Lambda e^{-2.68 R_J^2}, 
\label{eq:T_N}
\end{equation}
where $R_J = \jp/(|\jone|+|\jtwo|)$ is the ratio of the intrachain coupling to the interchain couplings (see Table \ref{tab:DFT_results} and Figure \ref{fig:J_labels}) and $\Lambda = 24.27\jp/k_B$ is a non-universal parameter.\cite{Barzykin2001} Evaluating Eq. \ref{eq:T_N} for the unbent \cuacac\ crystal gives $T_N\approx 1\times10^{-33}$\,K. Thus, we predict that the unbent crystal of \cuacac\ will be magnetically disordered down to the lowest experimentally reachable temperatures -- experimentally \cuacac\ will appear as an almost perfect TLL. To highlight the extreme one-dimensionality of \cuacac\ compared to other materials, one can make the striking comparison of $T_N/\jp\approx 10^{-33}$ to the other materials in Table \ref{tab:other_tns}.

\begin{figure*}
	\centering
	\includegraphics[width=\textwidth]{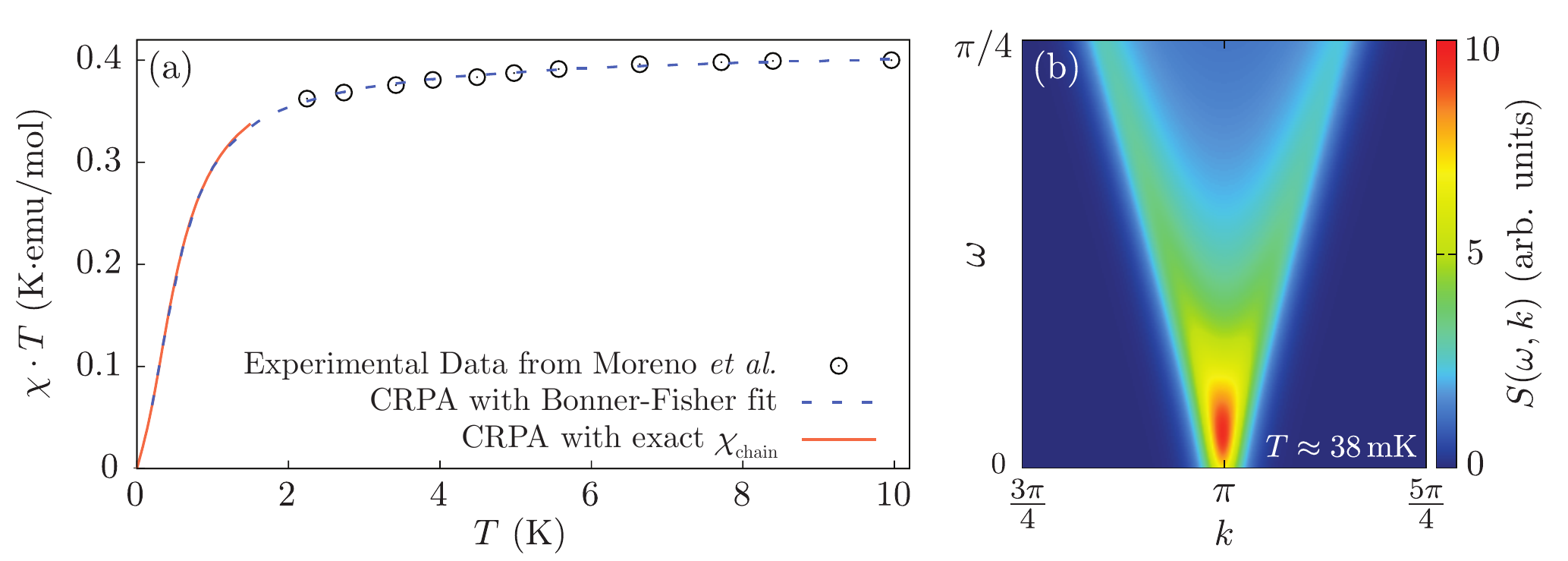}
	\caption{We predict that the experimental properties of \cuacac\ will closely mimic an isolated spin-1/2 Heisenberg chain, being an almost perfect TLL. (a) A fit of the CRPA with the 1D Bonner-Fisher susceptibility (Eq. S.17 with $\jp=0.75$\,K and $\jone+\jtwo=0.14$\,K) to experimental bulk susceptibility data from Moreno \textit{et al.} \cite{Moreno2013} and a low-temperature prediction with the CRPA and the exact 1D calculation from Eggert \textit{et al.}\cite{Eggert1994} (Eq. S.18). (b) Calculated plot of the dynamical structure factor of \cuacac\ with Eq. S.19. The 1D model of the bulk susceptibility is very successful and the dynamical structure factor prediction shows little deviation from an isolated 1D chain.}
	\label{fig:experiments}
\end{figure*}

Given the form of Eq. \ref{eq:T_N}, it is clear that $T_N$ is very sensitive to  $R_J$; because $T_N$ decays exponentially as a function of $R_J^2$, a small change in $R_J$ leads to a dramatic change in $T_N$. This extreme sensitivity is caused by the geometry of \cuacac; the interchain interactions are geometrically frustrated, as illustrated in Fig. \ref{fig:J_labels}. It is instructive to compare the $T_N$ calculated above with that of an unfrustrated analogue -- a cubic lattice where $\jone$ and $\jtwo$ are the same magnitude as in \cuacac, but their directions are perpendicular to $\jp$ (see Fig. \ref{fig:frustration}a).  The same CRPA calculation as above then results in 
\begin{equation}
T_N^\mathrm{cubic}\approx 0.56\frac{\jp}{k_B R_J}\sqrt{\log\left(\frac{\Lambda}{T_N}\right)},
\label{eq:cubic_tn}
\end{equation}
which yields $T_N^\mathrm{cubic}\approx 0.23\jp/k_B\approx 0.17\,\mathrm{K}$ using the parameters in Table \ref{tab:DFT_results}. This is 32 orders of magnitude higher than $T_N$ for the frustrated \cuacac\ lattice. Moreover, in contrast to the exponential dependence of Eq. \ref{eq:T_N}, $T_N^\mathrm{cubic}$ is proportional to $1/R_J$ -- it is larger and less sensitive to small changes in the value of $R_J$. This will be important when we discuss the bent crystals. 

The large contrast between geometrically frustrated and unfrustrated interactions is also demonstrated in previous work on the 2D anisotropic triangular lattice Heisenberg model, for Cs$_2$CuCl$_4$ in particular.\cite{Starykh2010, Bocquet2001}

We predict that the experimental properties of \cuacac\ will closely mimic an isolated spin-1/2 Heisenberg chain, displaying properties of an almost perfect TLL. The CRPA prediction of the bulk magnetic susceptibility using the exact 1D theory is limited to the low-temperature regime studied by Eggert \textit{et al}. \cite{Eggert1994} Conversely, the bulk susceptibility of \cuacac\ has only been measured above 2\,K, with no magnetic ordering detected. \cite{Moreno2013} Therefore, to compare our prediction with experiment, we first fit the CRPA using the Bonner-Fisher susceptibility of a single spin chain, which is successful in other materials at higher temperatures.\cite{Bonner1964, Estes1977, Ami1995} We set $\jp=0.75$\,K (our BS-DFT result) and found that the best fit corresponded to $\jone+\jtwo=0.14$\,K, in reasonable agreement our BS-DFT results for the interchain couplings. We used this value of $\jone+\jtwo$ to parametrize our low-temperature prediction. More details of the fit are given in the Supplementary Information. Fig. \ref{fig:experiments}(a) shows the Bonner-Fisher fit to the experimental data from Moreno \textit{et al.}\cite{Moreno2013} and our prediction of the low temperature magnetic susceptibility. The agreement is exceptional.

The dynamical structure factor for \cuacac\ has not been measured. Our dynamical structure factor prediction in Fig. \ref{fig:experiments}(b) (measurable via neutron scattering experiments) was calculated with our BS-DFT exchange parameters (Table \ref{tab:DFT_results}). It shows a slight asymmetry, which is absent for a TLL in an isolated Heisenberg chain at low temperatures. There are no adjustable parameters in this prediction.

\begin{figure}
\centering
\includegraphics[width=8cm]{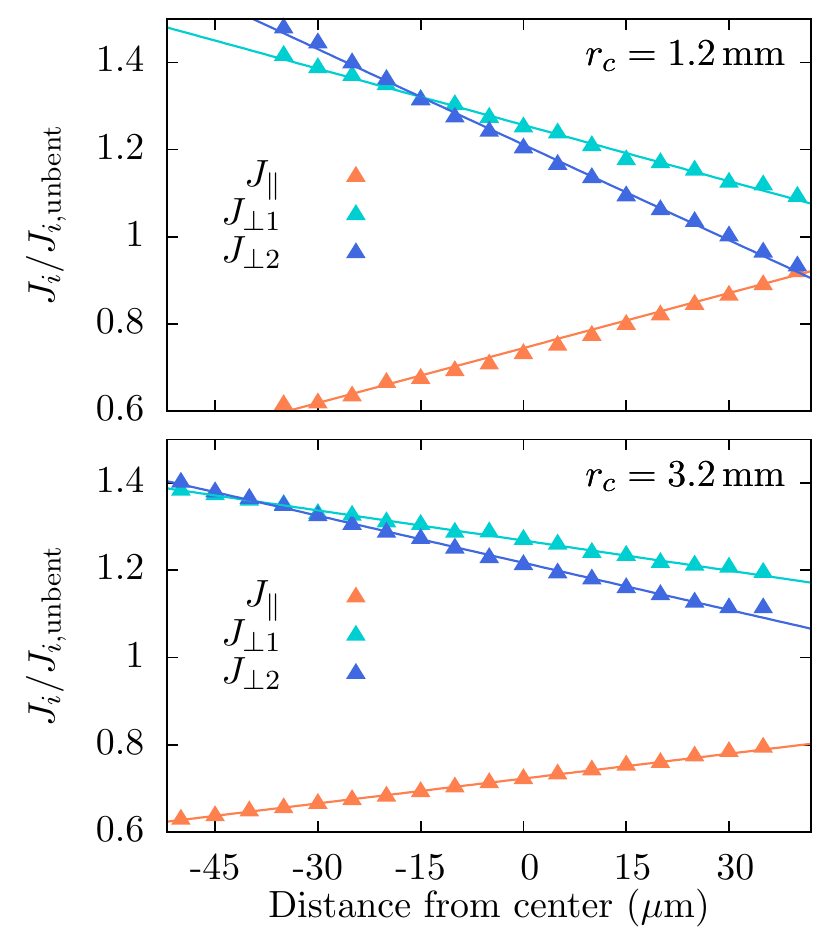}
\caption{BS-DFT calculations of the magnetic interactions in bent \cuacac\ using crystal data across bent samples, with different radii of curvature, $r_c$, from Worthy \textit{et al.} \cite{Worthy2017} The intra and interchain exchange couplings change as a function of the distance across a bent sample of \cuacac.  The center is defined as the position where the magnitude of the crystallographic $b$-axis is most similar to that of the  unbent structure (although, note that the $a$ and $c$ axes are quite different). Lines are a guide to the eye.}
\label{fig:Js}
\end{figure}

\textit{Bent Crystal.} Our BS-DFT results across the bent crystals of \cuacac\ are shown in Figure \ref{fig:Js}, where we plot $\jp$, $\jone$, and $\jtwo$ as ratios of the unbent parameters across each of the bent crystals. We calculate that the interchain coupling changes by over 20\% as a consequence of the crystal distortion. 

\begin{figure*}
\centering
\includegraphics[width=\textwidth]{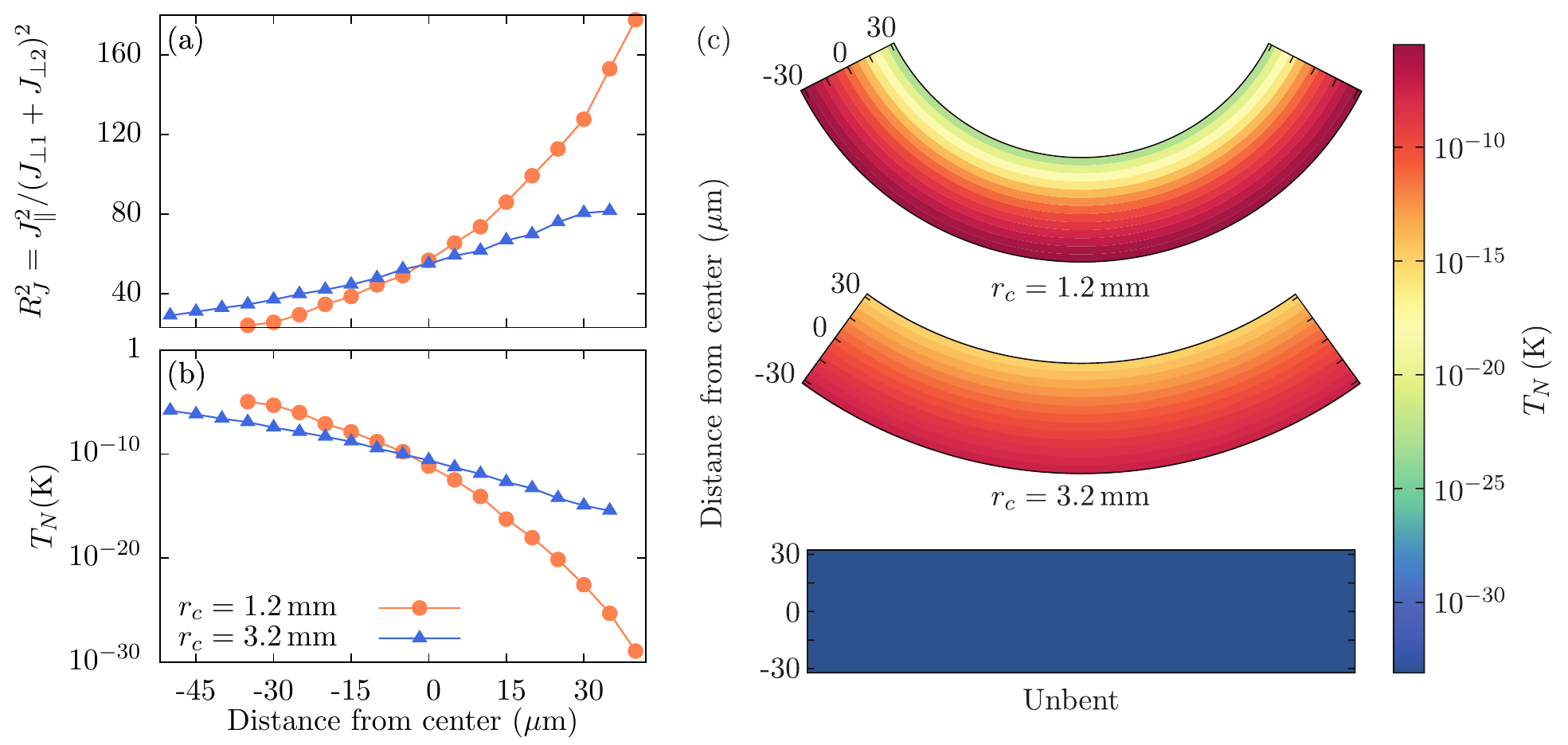}
\caption{There is a significant change in the square of the ratio of intra to interchain exchange coupling $R_J^2$ across both bends, (a), leading to a drastic change in N\'eel temperature $T_N$, (b) and (c). Note the logarithmic scale of the ordinate in panel (b). The center is defined as the position where the magnitude of the crystallographic $b$-axis is most similar to that of the  unbent structure (note, however, that the $a$ and $c$ axes are quite different).}
\label{fig:Tns}
\end{figure*}

Figure \ref{fig:Tns} shows $R_J^2$ and $T_N$ across the bent crystals. The change in geometry brought on by bending the \cuacac\ crystals causes a significant change in magnetic behaviour at different points across the bend; a small change in $R_J^2$ causes a very large change in the ordering temperature. In the most bent crystal, this means a change in $T_N$ of 24 orders of magnitude from one side of the bend to the other.

\begin{figure}
\centering
\includegraphics[width=7cm]{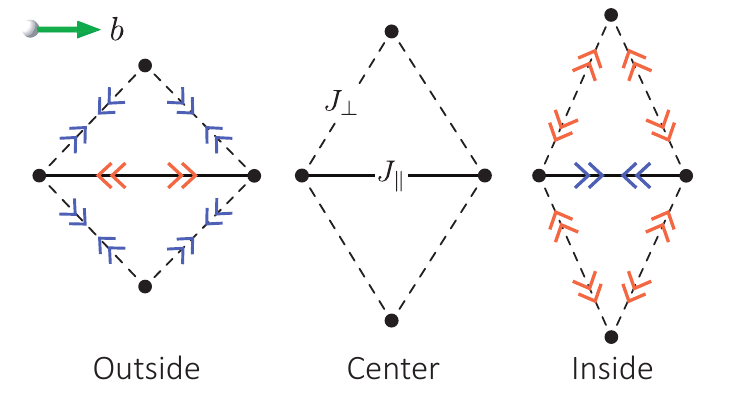}
\caption{When \cuacac\ is bent, the $b$-axis is stretched on the outside of the bend relative to the center part of the crystal (decreasing $\jp$), and compressed on the inside (increasing $\jp$). Conversely, the interchain separation decreases on the outside (increasing $\jone$ and $\jtwo$), and increases on the inside (decreasing $\jone$ and $\jtwo$). This leads to a dramatic increase of $T_N$ on the outside and a dramatic decrease of $T_N$ on the inside of the bend compared to the centre. This happens in both planes containing the chain ((a) and (b) in Fig. \ref{fig:J_labels})}
\label{fig:bend_changes}
\end{figure}

When the lattice is strained by bending, this causes a simultaneous, but opposite, change in $\jp$ and the perpendicular couplings, $\jone$ and $\jtwo$, relative to the center of the bent crystal -- illustrated in Fig. \ref{fig:bend_changes}. On the inside of the bend, the distance between copper atoms along the chain is smaller than in the center, due to the compression of the lattice along the $b$-axis, leading to a relative increase in $\jp$. Whereas, the distance between the chains \textit{increases} because the lattice is expanded along the $a$ and $c$ axes relative to the center, \textit{decreasing} $\jone$ and $\jtwo$. Both of these processes independently decrease $T_N$. On the outside of the bend, the opposite effect occurs; the $b$-axis is elongated causing $\jp$ to decrease and the $a$ and $c$ axes are compressed causing the interchain couplings to increase, leading to an increase in $T_N$. 

The transition at $T_N$ is an antiferromagnetic transition, which could be detected with the divergence of the magnetic susceptibility. We predict that, in a bent crystal, the bulk magnetic susceptibility would be a superposition of single chain transitions resulting from the different $T_N$ values at different points across the crystal.

Geometric frustration plays a vital role in this dramatic change in $T_N$ across the bend; the extreme sensitivity of $T_N$ to the changes in the crystal described above is due to the exponential dependence of $T_N$ on $R_J$ (Eq. \ref{eq:T_N}). If the lattice was cubic, the N\'eel temperature would have stronger proportionality to $R_J$ (Eq. \ref{eq:cubic_tn}), and one would not observe such a dramatic change in $T_N$ (also, the unbent crystal would have a much larger $T_N$).

\section*{Conclusions}

In conclusion, we predict that the magnetic ordering temperature of elastically flexible \cuacac\ changes dramatically when the material is bent. The unbent crystal will behave, experimentally, like an almost perfect TLL (i.e. uncoupled 1D spin chains). When the sample is bent perpendicular to the chain direction, the crystal geometry changes in such a way to maximally affect the value of the N\'eel temperature, $T_N$. A stretched crystal with a change in exchange couplings of 20\% has a theoretical ordering temperature of 0.01\,mK, which is 24 orders of magnitude higher than the unbent crystal, with a N\'eel temperature of $\sim10^{-33}$\,K. This change in $T_N$ across a bend would be experimentally evidenced by measuring the bulk susceptibility. The interchain interactions only weakly renormalize the properties of \cuacac\ relative to a single Heisenberg chain. This is due in part to the weakness of the interchain couplings, $\jone$ and $\jtwo$, but mostly to the presence of geometric frustration in the lattice; geometric frustration leads to the exponential suppression of the N\'eel temperature, stabilizing the Tomonaga-Luttinger spin-liquid phase. Our results provide a powerful proof-of-principle demonstration that magnetic interactions can be controlled via bending flexible crystals. We have demonstrated the possibility of using elastic flexible magnetic crystals to passively sense small deformations, curvatures, or flexures with extremely high precision by detecting the divergence of the magnetic susceptibility in the sample.

\cuacac\ has a N\'eel temperature that is highly sensitive to bending, but its extreme geometric frustration means that $k_BT_N$ is many orders of magnitude smaller than the magnetic exchange interactions. Therefore, simply increasing the exchange couplings would not be expected to lead to experimentally accessible N\'eel temperatures. Rather, as highlighted in Table \ref{tab:other_tns}, the extreme geometrical frustration of \cuacac\ is actually responsible for the low $T_N$. This suggests that an incompletely frustrated material may open the door to mechanomagnetics at experimentally accessible temperatures. However, our results show that there is a trade-off. Geometrical frustration also enhances the sensitivity of $T_N$ to bending. Unfrustrated coupling leads to a higher, measurable $T_N$ but lowers its sensitivity. Therefore, partial geometrical frustration, e.g., imperfectly triangular couplings perpendicular to the chain, could provide a balance with both a high $T_N$ and a strong sensitivity to bending.  We hope that this insight will play a key role in the future search and design of elastically flexible mechanomagnetic crystals.

\section*{Acknowledgement}

The authors thank Jack Clegg, Arnaud Grosjean, Amie Khosla, and Ross McKenzie for helpful conversations. We are indebted to Sebastian Eggert for making the data from Ref.~\citenum{Eggert1994} available at Ref.~\citenum{EggertDat}.
This work was supported by the Australian Research Council through Grants No. FT130100161 and DP160100060.

\bibliography{cu_acac}

\begin{thebibliography}{67}%
\makeatletter
\providecommand \@ifxundefined [1]{%
 \@ifx{#1\undefined}
}%
\providecommand \@ifnum [1]{%
 \ifnum #1\expandafter \@firstoftwo
 \else \expandafter \@secondoftwo
 \fi
}%
\providecommand \@ifx [1]{%
 \ifx #1\expandafter \@firstoftwo
 \else \expandafter \@secondoftwo
 \fi
}%
\providecommand \natexlab [1]{#1}%
\providecommand \enquote  [1]{``#1''}%
\providecommand \bibnamefont  [1]{#1}%
\providecommand \bibfnamefont [1]{#1}%
\providecommand \citenamefont [1]{#1}%
\providecommand \href@noop [0]{\@secondoftwo}%
\providecommand \href [0]{\begingroup \@sanitize@url \@href}%
\providecommand \@href[1]{\@@startlink{#1}\@@href}%
\providecommand \@@href[1]{\endgroup#1\@@endlink}%
\providecommand \@sanitize@url [0]{\catcode `\\12\catcode `\$12\catcode
  `\&12\catcode `\#12\catcode `\^12\catcode `\_12\catcode `\%12\relax}%
\providecommand \@@startlink[1]{}%
\providecommand \@@endlink[0]{}%
\providecommand \url  [0]{\begingroup\@sanitize@url \@url }%
\providecommand \@url [1]{\endgroup\@href {#1}{\urlprefix }}%
\providecommand \urlprefix  [0]{URL }%
\providecommand \Eprint [0]{\href }%
\providecommand \doibase [0]{http://dx.doi.org/}%
\providecommand \selectlanguage [0]{\@gobble}%
\providecommand \bibinfo  [0]{\@secondoftwo}%
\providecommand \bibfield  [0]{\@secondoftwo}%
\providecommand \translation [1]{[#1]}%
\providecommand \BibitemOpen [0]{}%
\providecommand \bibitemStop [0]{}%
\providecommand \bibitemNoStop [0]{.\EOS\space}%
\providecommand \EOS [0]{\spacefactor3000\relax}%
\providecommand \BibitemShut  [1]{\csname bibitem#1\endcsname}%
\let\auto@bib@innerbib\@empty
\bibitem [{\citenamefont {Chen}\ \emph {et~al.}(2014)\citenamefont {Chen},
  \citenamefont {Ghosh}, \citenamefont {Malla~Reddy},\ and\ \citenamefont
  {Buehler}}]{Chen2014}%
  \BibitemOpen
  \bibfield  {author} {\bibinfo {author} {\bibfnamefont {C.-T.}\ \bibnamefont
  {Chen}}, \bibinfo {author} {\bibfnamefont {S.}~\bibnamefont {Ghosh}},
  \bibinfo {author} {\bibfnamefont {C.}~\bibnamefont {Malla~Reddy}}, \ and\
  \bibinfo {author} {\bibfnamefont {M.~J.}\ \bibnamefont {Buehler}},\ }\href
  {\doibase 10.1039/C3CP55117B} {\bibfield  {journal} {\bibinfo  {journal}
  {Phys. Chem. Chem. Phys.}\ }\textbf {\bibinfo {volume} {16}},\ \bibinfo
  {pages} {13165} (\bibinfo {year} {2014})}\BibitemShut {NoStop}%
\bibitem [{\citenamefont {Ghosh}\ \emph
  {et~al.}(2015{\natexlab{a}})\citenamefont {Ghosh}, \citenamefont {Mishra},
  \citenamefont {Ganguly},\ and\ \citenamefont {Desiraju}}]{Ghosh2015jacs}%
  \BibitemOpen
  \bibfield  {author} {\bibinfo {author} {\bibfnamefont {S.}~\bibnamefont
  {Ghosh}}, \bibinfo {author} {\bibfnamefont {M.~K.}\ \bibnamefont {Mishra}},
  \bibinfo {author} {\bibfnamefont {S.}~\bibnamefont {Ganguly}}, \ and\
  \bibinfo {author} {\bibfnamefont {G.~R.}\ \bibnamefont {Desiraju}},\ }\href
  {\doibase 10.1021/jacs.5b05324} {\bibfield  {journal} {\bibinfo  {journal}
  {JACS}\ }\textbf {\bibinfo {volume} {137}},\ \bibinfo {pages} {9912}
  (\bibinfo {year} {2015}{\natexlab{a}})}\BibitemShut {NoStop}%
\bibitem [{\citenamefont {Ghosh}\ \emph
  {et~al.}(2015{\natexlab{b}})\citenamefont {Ghosh}, \citenamefont {Mishra},
  \citenamefont {Kadambi}, \citenamefont {Ramamurty},\ and\ \citenamefont
  {Desiraju}}]{Ghosh2015angew}%
  \BibitemOpen
  \bibfield  {author} {\bibinfo {author} {\bibfnamefont {S.}~\bibnamefont
  {Ghosh}}, \bibinfo {author} {\bibfnamefont {M.~K.}\ \bibnamefont {Mishra}},
  \bibinfo {author} {\bibfnamefont {S.~B.}\ \bibnamefont {Kadambi}}, \bibinfo
  {author} {\bibfnamefont {U.}~\bibnamefont {Ramamurty}}, \ and\ \bibinfo
  {author} {\bibfnamefont {G.~R.}\ \bibnamefont {Desiraju}},\ }\href {\doibase
  10.1002/anie.201410730} {\bibfield  {journal} {\bibinfo  {journal} {Angew.
  Chem. Int. Ed.}\ }\textbf {\bibinfo {volume} {54}},\ \bibinfo {pages} {2674}
  (\bibinfo {year} {2015}{\natexlab{b}})}\BibitemShut {NoStop}%
\bibitem [{\citenamefont {Hayashi}\ and\ \citenamefont
  {Koizumi}(2016)}]{Hayashi2016}%
  \BibitemOpen
  \bibfield  {author} {\bibinfo {author} {\bibfnamefont {S.}~\bibnamefont
  {Hayashi}}\ and\ \bibinfo {author} {\bibfnamefont {T.}~\bibnamefont
  {Koizumi}},\ }\href {\doibase 10.1002/ange.201509319} {\bibfield  {journal}
  {\bibinfo  {journal} {Angew. Chem.}\ }\textbf {\bibinfo {volume} {128}},\
  \bibinfo {pages} {2751} (\bibinfo {year} {2016})}\BibitemShut {NoStop}%
\bibitem [{\citenamefont {Worthy}\ \emph {et~al.}(2018)\citenamefont {Worthy},
  \citenamefont {Grosjean}, \citenamefont {Pfrunder}, \citenamefont {Xu},
  \citenamefont {Yan}, \citenamefont {Edwards}, \citenamefont {Clegg},\ and\
  \citenamefont {McMurtrie}}]{Worthy2017}%
  \BibitemOpen
  \bibfield  {author} {\bibinfo {author} {\bibfnamefont {A.}~\bibnamefont
  {Worthy}}, \bibinfo {author} {\bibfnamefont {A.}~\bibnamefont {Grosjean}},
  \bibinfo {author} {\bibfnamefont {M.~C.}\ \bibnamefont {Pfrunder}}, \bibinfo
  {author} {\bibfnamefont {Y.}~\bibnamefont {Xu}}, \bibinfo {author}
  {\bibfnamefont {C.}~\bibnamefont {Yan}}, \bibinfo {author} {\bibfnamefont
  {G.}~\bibnamefont {Edwards}}, \bibinfo {author} {\bibfnamefont {J.~K.}\
  \bibnamefont {Clegg}}, \ and\ \bibinfo {author} {\bibfnamefont {J.~C.}\
  \bibnamefont {McMurtrie}},\ }\href {\doibase 10.1038/nchem.2848} {\bibfield
  {journal} {\bibinfo  {journal} {Nat. Chem.}\ }\textbf {\bibinfo {volume}
  {10}},\ \bibinfo {pages} {65} (\bibinfo {year} {2018})}\BibitemShut {NoStop}%
\bibitem [{\citenamefont {Hayashi}\ and\ \citenamefont
  {Koizumi}(2018)}]{Hayashi2018}%
  \BibitemOpen
  \bibfield  {author} {\bibinfo {author} {\bibfnamefont {S.}~\bibnamefont
  {Hayashi}}\ and\ \bibinfo {author} {\bibfnamefont {T.}~\bibnamefont
  {Koizumi}},\ }\href {\doibase 10.1002/chem.201801965} {\bibfield  {journal}
  {\bibinfo  {journal} {Chem. Eur. J.}\ }\textbf {\bibinfo {volume} {24}},\
  \bibinfo {pages} {8507} (\bibinfo {year} {2018})}\BibitemShut {NoStop}%
\bibitem [{\citenamefont {Ahmed}\ \emph {et~al.}(2018)\citenamefont {Ahmed},
  \citenamefont {Karothu},\ and\ \citenamefont {Naumov}}]{Ahmed2018}%
  \BibitemOpen
  \bibfield  {author} {\bibinfo {author} {\bibfnamefont {E.}~\bibnamefont
  {Ahmed}}, \bibinfo {author} {\bibfnamefont {D.~P.}\ \bibnamefont {Karothu}},
  \ and\ \bibinfo {author} {\bibfnamefont {P.}~\bibnamefont {Naumov}},\ }\href
  {\doibase 10.1002/anie.201800137} {\bibfield  {journal} {\bibinfo  {journal}
  {Angew. Chem. Int. Ed.}\ }\textbf {\bibinfo {volume} {57}},\ \bibinfo {pages}
  {8837} (\bibinfo {year} {2018})}\BibitemShut {NoStop}%
\bibitem [{\citenamefont {Reddy}\ \emph {et~al.}(2006)\citenamefont {Reddy},
  \citenamefont {Padmanabhan},\ and\ \citenamefont {Desiraju}}]{Reddy2006}%
  \BibitemOpen
  \bibfield  {author} {\bibinfo {author} {\bibfnamefont {C.~M.}\ \bibnamefont
  {Reddy}}, \bibinfo {author} {\bibfnamefont {K.~A.}\ \bibnamefont
  {Padmanabhan}}, \ and\ \bibinfo {author} {\bibfnamefont {G.~R.}\ \bibnamefont
  {Desiraju}},\ }\href {\doibase 10.1021/cg060398w} {\bibfield  {journal}
  {\bibinfo  {journal} {Cryst. Growth Des.}\ }\textbf {\bibinfo {volume} {6}},\
  \bibinfo {pages} {2720} (\bibinfo {year} {2006})}\BibitemShut {NoStop}%
\bibitem [{\citenamefont {Mishra}\ \emph {et~al.}(2016)\citenamefont {Mishra},
  \citenamefont {Ramamurty},\ and\ \citenamefont {Desiraju}}]{MishraReview}%
  \BibitemOpen
  \bibfield  {author} {\bibinfo {author} {\bibfnamefont {M.~K.}\ \bibnamefont
  {Mishra}}, \bibinfo {author} {\bibfnamefont {U.}~\bibnamefont {Ramamurty}}, \
  and\ \bibinfo {author} {\bibfnamefont {G.~R.}\ \bibnamefont {Desiraju}},\
  }\href {\doibase 10.1016/j.cossms.2016.05.011} {\bibfield  {journal}
  {\bibinfo  {journal} {Curr. Opin. Solid State Mater. Sci.}\ }\textbf
  {\bibinfo {volume} {20}},\ \bibinfo {pages} {361} (\bibinfo {year}
  {2016})}\BibitemShut {NoStop}%
\bibitem [{\citenamefont {Saha}\ \emph {et~al.}(2018)\citenamefont {Saha},
  \citenamefont {Mishra}, \citenamefont {Reddy},\ and\ \citenamefont
  {Desiraju}}]{Saha2018}%
  \BibitemOpen
  \bibfield  {author} {\bibinfo {author} {\bibfnamefont {S.}~\bibnamefont
  {Saha}}, \bibinfo {author} {\bibfnamefont {M.~K.}\ \bibnamefont {Mishra}},
  \bibinfo {author} {\bibfnamefont {C.~M.}\ \bibnamefont {Reddy}}, \ and\
  \bibinfo {author} {\bibfnamefont {G.~R.}\ \bibnamefont {Desiraju}},\ }\href
  {\doibase 10.1021/acs.accounts.8b00425} {\bibfield  {journal} {\bibinfo
  {journal} {Acc. Chem. Res.}\ }\textbf {\bibinfo {volume} {51}},\ \bibinfo
  {pages} {2957} (\bibinfo {year} {2018})}\BibitemShut {NoStop}%
\bibitem [{\citenamefont {Brock}\ \emph {et~al.}(2018)\citenamefont {Brock},
  \citenamefont {Whittaker}, \citenamefont {Powell}, \citenamefont {Pfrunder},
  \citenamefont {Grosjean}, \citenamefont {Parsons}, \citenamefont
  {McMurtrie},\ and\ \citenamefont {Clegg}}]{Brock2018}%
  \BibitemOpen
  \bibfield  {author} {\bibinfo {author} {\bibfnamefont {A.~J.}\ \bibnamefont
  {Brock}}, \bibinfo {author} {\bibfnamefont {J.~J.}\ \bibnamefont
  {Whittaker}}, \bibinfo {author} {\bibfnamefont {J.~A.}\ \bibnamefont
  {Powell}}, \bibinfo {author} {\bibfnamefont {M.~C.}\ \bibnamefont
  {Pfrunder}}, \bibinfo {author} {\bibfnamefont {A.}~\bibnamefont {Grosjean}},
  \bibinfo {author} {\bibfnamefont {S.}~\bibnamefont {Parsons}}, \bibinfo
  {author} {\bibfnamefont {J.~C.}\ \bibnamefont {McMurtrie}}, \ and\ \bibinfo
  {author} {\bibfnamefont {J.~K.}\ \bibnamefont {Clegg}},\ }\href {\doibase
  10.1002/anie.201806431} {\bibfield  {journal} {\bibinfo  {journal} {Angew.
  Chem. Int. Ed.}\ }\textbf {\bibinfo {volume} {57}},\ \bibinfo {pages} {11325}
  (\bibinfo {year} {2018})}\BibitemShut {NoStop}%
\bibitem [{\citenamefont {Zakharov}\ and\ \citenamefont
  {Boldyreva}(2019)}]{Zakharov2019}%
  \BibitemOpen
  \bibfield  {author} {\bibinfo {author} {\bibfnamefont {B.~A.}\ \bibnamefont
  {Zakharov}}\ and\ \bibinfo {author} {\bibfnamefont {E.~V.}\ \bibnamefont
  {Boldyreva}},\ }\href {\doibase 10.1039/C8CE01391H} {\bibfield  {journal}
  {\bibinfo  {journal} {CrystEngComm}\ } (\bibinfo {year} {2019}),\
  10.1039/C8CE01391H}\BibitemShut {NoStop}%
\bibitem [{\citenamefont {Hayashi}\ \emph {et~al.}(2017)\citenamefont
  {Hayashi}, \citenamefont {Koizumi},\ and\ \citenamefont
  {Kamiya}}]{Hayashi2017}%
  \BibitemOpen
  \bibfield  {author} {\bibinfo {author} {\bibfnamefont {S.}~\bibnamefont
  {Hayashi}}, \bibinfo {author} {\bibfnamefont {T.}~\bibnamefont {Koizumi}}, \
  and\ \bibinfo {author} {\bibfnamefont {N.}~\bibnamefont {Kamiya}},\ }\href
  {\doibase 10.1021/acs.cgd.7b00992} {\bibfield  {journal} {\bibinfo  {journal}
  {Crystal Growth \& Design}\ }\textbf {\bibinfo {volume} {17}},\ \bibinfo
  {pages} {6158} (\bibinfo {year} {2017})}\BibitemShut {NoStop}%
\bibitem [{\citenamefont {Hayashi}\ \emph {et~al.}(2018)\citenamefont
  {Hayashi}, \citenamefont {Yamamoto}, \citenamefont {Takeuchi}, \citenamefont
  {Ie},\ and\ \citenamefont {Takagi}}]{Hayashi2018angew}%
  \BibitemOpen
  \bibfield  {author} {\bibinfo {author} {\bibfnamefont {S.}~\bibnamefont
  {Hayashi}}, \bibinfo {author} {\bibfnamefont {S.-y.}\ \bibnamefont
  {Yamamoto}}, \bibinfo {author} {\bibfnamefont {D.}~\bibnamefont {Takeuchi}},
  \bibinfo {author} {\bibfnamefont {Y.}~\bibnamefont {Ie}}, \ and\ \bibinfo
  {author} {\bibfnamefont {K.}~\bibnamefont {Takagi}},\ }\href {\doibase
  10.1002/anie.201810422} {\bibfield  {journal} {\bibinfo  {journal} {Angew.
  Chem. Int. Ed.}\ }\textbf {\bibinfo {volume} {57}},\ \bibinfo {pages} {17002}
  (\bibinfo {year} {2018})}\BibitemShut {NoStop}%
\bibitem [{\citenamefont {Saggio}\ \emph {et~al.}(2016)\citenamefont {Saggio},
  \citenamefont {Riillo}, \citenamefont {Sbernini},\ and\ \citenamefont
  {Quitadamo}}]{Saggio2016}%
  \BibitemOpen
  \bibfield  {author} {\bibinfo {author} {\bibfnamefont {G.}~\bibnamefont
  {Saggio}}, \bibinfo {author} {\bibfnamefont {F.}~\bibnamefont {Riillo}},
  \bibinfo {author} {\bibfnamefont {L.}~\bibnamefont {Sbernini}}, \ and\
  \bibinfo {author} {\bibfnamefont {L.~R.}\ \bibnamefont {Quitadamo}},\
  }\href@noop {} {\bibfield  {journal} {\bibinfo  {journal} {Smart Mater.
  Struct.}\ }\textbf {\bibinfo {volume} {25}},\ \bibinfo {pages} {013001}
  (\bibinfo {year} {2016})}\BibitemShut {NoStop}%
\bibitem [{\citenamefont {Balents}(2010)}]{Balents2010}%
  \BibitemOpen
  \bibfield  {author} {\bibinfo {author} {\bibfnamefont {L.}~\bibnamefont
  {Balents}},\ }\href {\doibase 10.1038/nature08917} {\bibfield  {journal}
  {\bibinfo  {journal} {Nature}\ }\textbf {\bibinfo {volume} {464}},\ \bibinfo
  {pages} {199} (\bibinfo {year} {2010})}\BibitemShut {NoStop}%
\bibitem [{\citenamefont {Savary}\ and\ \citenamefont
  {Balents}(2017)}]{BalentsReview}%
  \BibitemOpen
  \bibfield  {author} {\bibinfo {author} {\bibfnamefont {L.}~\bibnamefont
  {Savary}}\ and\ \bibinfo {author} {\bibfnamefont {L.}~\bibnamefont
  {Balents}},\ }\href {\doibase 10.1088/0034-4885/80/1/016502} {\bibfield
  {journal} {\bibinfo  {journal} {Rep. Prog. Phys.}\ }\textbf {\bibinfo
  {volume} {80}},\ \bibinfo {pages} {016502} (\bibinfo {year}
  {2017})}\BibitemShut {NoStop}%
\bibitem [{\citenamefont {Tomonaga}(1950)}]{Tomonaga1950}%
  \BibitemOpen
  \bibfield  {author} {\bibinfo {author} {\bibfnamefont {S.}~\bibnamefont
  {Tomonaga}},\ }\href@noop {} {\bibfield  {journal} {\bibinfo  {journal}
  {Prog. Theor. Phys.}\ }\textbf {\bibinfo {volume} {5}},\ \bibinfo {pages}
  {544} (\bibinfo {year} {1950})}\BibitemShut {NoStop}%
\bibitem [{\citenamefont {Luttinger}(1963)}]{Luttinger1963}%
  \BibitemOpen
  \bibfield  {author} {\bibinfo {author} {\bibfnamefont {J.~M.}\ \bibnamefont
  {Luttinger}},\ }\href@noop {} {\bibfield  {journal} {\bibinfo  {journal} {J.
  Math. Phys.}\ }\textbf {\bibinfo {volume} {4}},\ \bibinfo {pages} {1154}
  (\bibinfo {year} {1963})}\BibitemShut {NoStop}%
\bibitem [{\citenamefont {Haldane}(1981)}]{Haldane1981}%
  \BibitemOpen
  \bibfield  {author} {\bibinfo {author} {\bibfnamefont {F.~D.~M.}\
  \bibnamefont {Haldane}},\ }\href@noop {} {\bibfield  {journal} {\bibinfo
  {journal} {J. Phys. C: Solid State Phys.}\ ,\ \bibinfo {pages} {2585}}
  (\bibinfo {year} {1981})}\BibitemShut {NoStop}%
\bibitem [{\citenamefont {Lake}\ \emph {et~al.}(2005)\citenamefont {Lake},
  \citenamefont {Tennant}, \citenamefont {Frost},\ and\ \citenamefont
  {Nagler}}]{Lake2005}%
  \BibitemOpen
  \bibfield  {author} {\bibinfo {author} {\bibfnamefont {B.}~\bibnamefont
  {Lake}}, \bibinfo {author} {\bibfnamefont {D.~A.}\ \bibnamefont {Tennant}},
  \bibinfo {author} {\bibfnamefont {C.~D.}\ \bibnamefont {Frost}}, \ and\
  \bibinfo {author} {\bibnamefont {Nagler}},\ }\href {\doibase
  10.1038/nmat1327} {\bibfield  {journal} {\bibinfo  {journal} {Nat. Mater.}\
  }\textbf {\bibinfo {volume} {4}},\ \bibinfo {pages} {329} (\bibinfo {year}
  {2005})}\BibitemShut {NoStop}%
\bibitem [{\citenamefont {Landee}\ and\ \citenamefont
  {Turnbull}(2013)}]{Landee2013}%
  \BibitemOpen
  \bibfield  {author} {\bibinfo {author} {\bibfnamefont {C.~P.}\ \bibnamefont
  {Landee}}\ and\ \bibinfo {author} {\bibfnamefont {M.~M.}\ \bibnamefont
  {Turnbull}},\ }\href {\doibase 10.1002/ejic.201300133} {\bibfield  {journal}
  {\bibinfo  {journal} {European Journal of Inorganic Chemistry}\ ,\ \bibinfo
  {pages} {2266}} (\bibinfo {year} {2013})}\BibitemShut {NoStop}%
\bibitem [{\citenamefont {Lancaster}\ \emph {et~al.}(2006)\citenamefont
  {Lancaster}, \citenamefont {Blundell}, \citenamefont {Brooks}, \citenamefont
  {Baker}, \citenamefont {Pratt}, \citenamefont {Manson}, \citenamefont
  {Landee},\ and\ \citenamefont {Baines}}]{Lancaster2006}%
  \BibitemOpen
  \bibfield  {author} {\bibinfo {author} {\bibfnamefont {T.}~\bibnamefont
  {Lancaster}}, \bibinfo {author} {\bibfnamefont {S.~J.}\ \bibnamefont
  {Blundell}}, \bibinfo {author} {\bibfnamefont {M.~L.}\ \bibnamefont
  {Brooks}}, \bibinfo {author} {\bibfnamefont {P.~J.}\ \bibnamefont {Baker}},
  \bibinfo {author} {\bibfnamefont {F.~L.}\ \bibnamefont {Pratt}}, \bibinfo
  {author} {\bibfnamefont {J.~L.}\ \bibnamefont {Manson}}, \bibinfo {author}
  {\bibfnamefont {C.~P.}\ \bibnamefont {Landee}}, \ and\ \bibinfo {author}
  {\bibfnamefont {C.}~\bibnamefont {Baines}},\ }\href@noop {} {\bibfield
  {journal} {\bibinfo  {journal} {Phys. Rev. B}\ }\textbf {\bibinfo {volume}
  {73}},\ \bibinfo {pages} {020410} (\bibinfo {year} {2006})}\BibitemShut
  {NoStop}%
\bibitem [{\citenamefont {Breunig}\ \emph {et~al.}(2017)\citenamefont
  {Breunig}, \citenamefont {Garst}, \citenamefont {Kl{\"u}mper}, \citenamefont
  {Rohrkamp}, \citenamefont {Turnbull},\ and\ \citenamefont
  {Lorenz}}]{Breunig2017}%
  \BibitemOpen
  \bibfield  {author} {\bibinfo {author} {\bibfnamefont {O.}~\bibnamefont
  {Breunig}}, \bibinfo {author} {\bibfnamefont {M.}~\bibnamefont {Garst}},
  \bibinfo {author} {\bibfnamefont {A.}~\bibnamefont {Kl{\"u}mper}}, \bibinfo
  {author} {\bibfnamefont {J.}~\bibnamefont {Rohrkamp}}, \bibinfo {author}
  {\bibfnamefont {M.~M.}\ \bibnamefont {Turnbull}}, \ and\ \bibinfo {author}
  {\bibfnamefont {T.}~\bibnamefont {Lorenz}},\ }\href {\doibase
  10.1126/sciadv.aao3773} {\bibfield  {journal} {\bibinfo  {journal} {Science
  Advances}\ }\textbf {\bibinfo {volume} {3}} (\bibinfo {year} {2017}),\
  10.1126/sciadv.aao3773}\BibitemShut {NoStop}%
\bibitem [{\citenamefont {Jornet-Somoza}\ \emph {et~al.}(2010)\citenamefont
  {Jornet-Somoza}, \citenamefont {Deumal}, \citenamefont {Robb}, \citenamefont
  {Landee}, \citenamefont {Turnbull}, \citenamefont {Feyerherm},\ and\
  \citenamefont {Novoa}}]{Somoza2010}%
  \BibitemOpen
  \bibfield  {author} {\bibinfo {author} {\bibfnamefont {J.}~\bibnamefont
  {Jornet-Somoza}}, \bibinfo {author} {\bibfnamefont {M.}~\bibnamefont
  {Deumal}}, \bibinfo {author} {\bibfnamefont {M.~A.}\ \bibnamefont {Robb}},
  \bibinfo {author} {\bibfnamefont {C.~P.}\ \bibnamefont {Landee}}, \bibinfo
  {author} {\bibfnamefont {M.~M.}\ \bibnamefont {Turnbull}}, \bibinfo {author}
  {\bibfnamefont {R.}~\bibnamefont {Feyerherm}}, \ and\ \bibinfo {author}
  {\bibfnamefont {J.~J.}\ \bibnamefont {Novoa}},\ }\href {\doibase
  10.1021/ic902139h} {\bibfield  {journal} {\bibinfo  {journal} {Inorg. Chem.}\
  }\textbf {\bibinfo {volume} {49}},\ \bibinfo {pages} {1750} (\bibinfo {year}
  {2010})}\BibitemShut {NoStop}%
\bibitem [{\citenamefont {Rule}\ \emph {et~al.}(2011)\citenamefont {Rule},
  \citenamefont {Reehuis}, \citenamefont {Gibson}, \citenamefont {Ouladdiaf},
  \citenamefont {Gutmann}, \citenamefont {Hoffmann}, \citenamefont {Gerischer},
  \citenamefont {Tennant}, \citenamefont {S\"ullow},\ and\ \citenamefont
  {Lang}}]{Rule2011}%
  \BibitemOpen
  \bibfield  {author} {\bibinfo {author} {\bibfnamefont {K.~C.}\ \bibnamefont
  {Rule}}, \bibinfo {author} {\bibfnamefont {M.}~\bibnamefont {Reehuis}},
  \bibinfo {author} {\bibfnamefont {M.~C.~R.}\ \bibnamefont {Gibson}}, \bibinfo
  {author} {\bibfnamefont {B.}~\bibnamefont {Ouladdiaf}}, \bibinfo {author}
  {\bibfnamefont {M.~J.}\ \bibnamefont {Gutmann}}, \bibinfo {author}
  {\bibfnamefont {J.-U.}\ \bibnamefont {Hoffmann}}, \bibinfo {author}
  {\bibfnamefont {S.}~\bibnamefont {Gerischer}}, \bibinfo {author}
  {\bibfnamefont {D.~A.}\ \bibnamefont {Tennant}}, \bibinfo {author}
  {\bibfnamefont {S.}~\bibnamefont {S\"ullow}}, \ and\ \bibinfo {author}
  {\bibfnamefont {M.}~\bibnamefont {Lang}},\ }\href {\doibase
  10.1103/PhysRevB.83.104401} {\bibfield  {journal} {\bibinfo  {journal} {Phys.
  Rev. B}\ }\textbf {\bibinfo {volume} {83}},\ \bibinfo {pages} {104401}
  (\bibinfo {year} {2011})}\BibitemShut {NoStop}%
\bibitem [{\citenamefont {Coldea}\ \emph {et~al.}(2001)\citenamefont {Coldea},
  \citenamefont {Tennant}, \citenamefont {Tsvelik},\ and\ \citenamefont
  {Tylczynski}}]{Coldea2001}%
  \BibitemOpen
  \bibfield  {author} {\bibinfo {author} {\bibfnamefont {R.}~\bibnamefont
  {Coldea}}, \bibinfo {author} {\bibfnamefont {D.~A.}\ \bibnamefont {Tennant}},
  \bibinfo {author} {\bibfnamefont {A.~M.}\ \bibnamefont {Tsvelik}}, \ and\
  \bibinfo {author} {\bibfnamefont {Z.}~\bibnamefont {Tylczynski}},\ }\href
  {\doibase 10.1103/PhysRevLett.86.1335} {\bibfield  {journal} {\bibinfo
  {journal} {Phys. Rev. Lett.}\ }\textbf {\bibinfo {volume} {86}},\ \bibinfo
  {pages} {1335} (\bibinfo {year} {2001})}\BibitemShut {NoStop}%
\bibitem [{\citenamefont {Satija}\ \emph {et~al.}(1980)\citenamefont {Satija},
  \citenamefont {Axe}, \citenamefont {Shirane}, \citenamefont {Yoshizawa},\
  and\ \citenamefont {Hirakawa}}]{Satija1980}%
  \BibitemOpen
  \bibfield  {author} {\bibinfo {author} {\bibfnamefont {S.~K.}\ \bibnamefont
  {Satija}}, \bibinfo {author} {\bibfnamefont {J.~D.}\ \bibnamefont {Axe}},
  \bibinfo {author} {\bibfnamefont {G.}~\bibnamefont {Shirane}}, \bibinfo
  {author} {\bibfnamefont {H.}~\bibnamefont {Yoshizawa}}, \ and\ \bibinfo
  {author} {\bibfnamefont {K.}~\bibnamefont {Hirakawa}},\ }\href {\doibase
  10.1103/PhysRevB.21.2001} {\bibfield  {journal} {\bibinfo  {journal} {Phys.
  Rev. B}\ }\textbf {\bibinfo {volume} {21}},\ \bibinfo {pages} {2001}
  (\bibinfo {year} {1980})}\BibitemShut {NoStop}%
\bibitem [{\citenamefont {Hutchings}\ \emph {et~al.}(1969)\citenamefont
  {Hutchings}, \citenamefont {Samuelsen}, \citenamefont {Shirane},\ and\
  \citenamefont {Hirakawa}}]{Hutchings1969}%
  \BibitemOpen
  \bibfield  {author} {\bibinfo {author} {\bibfnamefont {M.~T.}\ \bibnamefont
  {Hutchings}}, \bibinfo {author} {\bibfnamefont {E.~J.}\ \bibnamefont
  {Samuelsen}}, \bibinfo {author} {\bibfnamefont {G.}~\bibnamefont {Shirane}},
  \ and\ \bibinfo {author} {\bibfnamefont {K.}~\bibnamefont {Hirakawa}},\
  }\href {\doibase 10.1103/PhysRev.188.919} {\bibfield  {journal} {\bibinfo
  {journal} {Phys. Rev.}\ }\textbf {\bibinfo {volume} {188}},\ \bibinfo {pages}
  {919} (\bibinfo {year} {1969})}\BibitemShut {NoStop}%
\bibitem [{\citenamefont {Goddard}\ \emph {et~al.}(2012)\citenamefont
  {Goddard}, \citenamefont {Manson}, \citenamefont {Singleton}, \citenamefont
  {Franke}, \citenamefont {Lancaster}, \citenamefont {Steele}, \citenamefont
  {Blundell}, \citenamefont {Baines}, \citenamefont {Pratt}, \citenamefont
  {McDonald}, \citenamefont {Ayala-Valenzuela}, \citenamefont {Corbey},
  \citenamefont {Southerland}, \citenamefont {Sengupta},\ and\ \citenamefont
  {Schlueter}}]{Goddard2012}%
  \BibitemOpen
  \bibfield  {author} {\bibinfo {author} {\bibfnamefont {P.~A.}\ \bibnamefont
  {Goddard}}, \bibinfo {author} {\bibfnamefont {J.~L.}\ \bibnamefont {Manson}},
  \bibinfo {author} {\bibfnamefont {J.}~\bibnamefont {Singleton}}, \bibinfo
  {author} {\bibfnamefont {I.}~\bibnamefont {Franke}}, \bibinfo {author}
  {\bibfnamefont {T.}~\bibnamefont {Lancaster}}, \bibinfo {author}
  {\bibfnamefont {A.~J.}\ \bibnamefont {Steele}}, \bibinfo {author}
  {\bibfnamefont {S.~J.}\ \bibnamefont {Blundell}}, \bibinfo {author}
  {\bibfnamefont {C.}~\bibnamefont {Baines}}, \bibinfo {author} {\bibfnamefont
  {F.~L.}\ \bibnamefont {Pratt}}, \bibinfo {author} {\bibfnamefont {R.~D.}\
  \bibnamefont {McDonald}}, \bibinfo {author} {\bibfnamefont {O.~E.}\
  \bibnamefont {Ayala-Valenzuela}}, \bibinfo {author} {\bibfnamefont {J.~F.}\
  \bibnamefont {Corbey}}, \bibinfo {author} {\bibfnamefont {H.~I.}\
  \bibnamefont {Southerland}}, \bibinfo {author} {\bibfnamefont
  {P.}~\bibnamefont {Sengupta}}, \ and\ \bibinfo {author} {\bibfnamefont
  {J.~A.}\ \bibnamefont {Schlueter}},\ }\href {\doibase
  10.1103/PhysRevLett.108.077208} {\bibfield  {journal} {\bibinfo  {journal}
  {Phys. Rev. Lett.}\ }\textbf {\bibinfo {volume} {108}},\ \bibinfo {pages}
  {077208} (\bibinfo {year} {2012})}\BibitemShut {NoStop}%
\bibitem [{\citenamefont {Noodleman}(1981)}]{Noodleman1981}%
  \BibitemOpen
  \bibfield  {author} {\bibinfo {author} {\bibfnamefont {L.}~\bibnamefont
  {Noodleman}},\ }\href {\doibase 10.1063/1.440939} {\bibfield  {journal}
  {\bibinfo  {journal} {J. Chem. Phys.}\ }\textbf {\bibinfo {volume} {74}},\
  \bibinfo {pages} {5737} (\bibinfo {year} {1981})}\BibitemShut {NoStop}%
\bibitem [{\citenamefont {Mouesca}(2014)}]{Mouesca}%
  \BibitemOpen
  \bibfield  {author} {\bibinfo {author} {\bibfnamefont {J.-M.}\ \bibnamefont
  {Mouesca}},\ }in\ \href@noop {} {\emph {\bibinfo {booktitle}
  {Metallo-proteins: Methods and Protocols}}},\ \bibinfo {editor} {edited by\
  \bibinfo {editor} {\bibfnamefont {J.~C.}\ \bibnamefont {Fontecilla-Camps}}\
  and\ \bibinfo {editor} {\bibfnamefont {Y.}~\bibnamefont {Nicolet}}}\
  (\bibinfo  {publisher} {Humana Press},\ \bibinfo {year} {2014})\
  Chap.~\bibinfo {chapter} {15}, pp.\ \bibinfo {pages} {269--296}\BibitemShut
  {NoStop}%
\bibitem [{\citenamefont {Bocquet}\ \emph {et~al.}(2001)\citenamefont
  {Bocquet}, \citenamefont {Essler}, \citenamefont {Tsvelik},\ and\
  \citenamefont {Gogolin}}]{Bocquet2001}%
  \BibitemOpen
  \bibfield  {author} {\bibinfo {author} {\bibfnamefont {M.}~\bibnamefont
  {Bocquet}}, \bibinfo {author} {\bibfnamefont {F.~H.~L.}\ \bibnamefont
  {Essler}}, \bibinfo {author} {\bibfnamefont {A.~M.}\ \bibnamefont {Tsvelik}},
  \ and\ \bibinfo {author} {\bibfnamefont {A.~O.}\ \bibnamefont {Gogolin}},\
  }\href {\doibase 10.1103/PhysRevB.64.094425} {\bibfield  {journal} {\bibinfo
  {journal} {Phys. Rev. B}\ }\textbf {\bibinfo {volume} {64}},\ \bibinfo
  {pages} {094425} (\bibinfo {year} {2001})}\BibitemShut {NoStop}%
\bibitem [{\citenamefont {Yamaguchi}\ \emph {et~al.}(1988)\citenamefont
  {Yamaguchi}, \citenamefont {Jensen}, \citenamefont {Dorigo},\ and\
  \citenamefont {Houk}}]{Yamaguchi1988}%
  \BibitemOpen
  \bibfield  {author} {\bibinfo {author} {\bibfnamefont {K.}~\bibnamefont
  {Yamaguchi}}, \bibinfo {author} {\bibfnamefont {F.}~\bibnamefont {Jensen}},
  \bibinfo {author} {\bibfnamefont {A.}~\bibnamefont {Dorigo}}, \ and\ \bibinfo
  {author} {\bibfnamefont {K.}~\bibnamefont {Houk}},\ }\href {\doibase
  10.1016/0009-2614(88)80378-6} {\bibfield  {journal} {\bibinfo  {journal}
  {Chem. Phys. Lett.}\ }\textbf {\bibinfo {volume} {149}},\ \bibinfo {pages}
  {537} (\bibinfo {year} {1988})}\BibitemShut {NoStop}%
\bibitem [{\citenamefont {Frisch}\ \emph {et~al.}()\citenamefont {Frisch} \emph
  {et~al.}}]{g09}%
  \BibitemOpen
  \bibfield  {author} {\bibinfo {author} {\bibfnamefont {M.~J.}\ \bibnamefont
  {Frisch}} \emph {et~al.},\ }\href@noop {} {\enquote {\bibinfo {title}
  {Gaussian~09 {R}evision {E}.01},}\ }\bibinfo {note} {Gaussian Inc.
  Wallingford CT 2009}\BibitemShut {NoStop}%
\bibitem [{\citenamefont {Becke}(1993)}]{B3LYPa}%
  \BibitemOpen
  \bibfield  {author} {\bibinfo {author} {\bibfnamefont {A.~D.}\ \bibnamefont
  {Becke}},\ }\href {\doibase 10.1063/1.464913} {\bibfield  {journal} {\bibinfo
   {journal} {J. Chem. Phys.}\ }\textbf {\bibinfo {volume} {98}},\ \bibinfo
  {pages} {5648} (\bibinfo {year} {1993})}\BibitemShut {NoStop}%
\bibitem [{\citenamefont {Stephens}\ \emph {et~al.}(1994)\citenamefont
  {Stephens}, \citenamefont {Devlin}, \citenamefont {Chabalowski},\ and\
  \citenamefont {Frisch}}]{B3LYPb}%
  \BibitemOpen
  \bibfield  {author} {\bibinfo {author} {\bibfnamefont {P.~J.}\ \bibnamefont
  {Stephens}}, \bibinfo {author} {\bibfnamefont {F.~J.}\ \bibnamefont
  {Devlin}}, \bibinfo {author} {\bibfnamefont {C.~F.}\ \bibnamefont
  {Chabalowski}}, \ and\ \bibinfo {author} {\bibfnamefont {M.~J.}\ \bibnamefont
  {Frisch}},\ }\href {\doibase 10.1021/j100096a001} {\bibfield  {journal}
  {\bibinfo  {journal} {J. Phys. Chem.}\ }\textbf {\bibinfo {volume} {98}},\
  \bibinfo {pages} {11623} (\bibinfo {year} {1994})}\BibitemShut {NoStop}%
\bibitem [{\citenamefont {{Dunning Jr.}}\ and\ \citenamefont
  {Hay}(1977)}]{Dunning1977}%
  \BibitemOpen
  \bibfield  {author} {\bibinfo {author} {\bibfnamefont {T.~H.}\ \bibnamefont
  {{Dunning Jr.}}}\ and\ \bibinfo {author} {\bibfnamefont {P.~J.}\ \bibnamefont
  {Hay}},\ }\enquote {\bibinfo {title} {Methods of electronic structure
  theory},}\ \ (\bibinfo  {publisher} {Plenum Press},\ \bibinfo {year} {1977})\
  \bibinfo {edition} {3rd}\ ed.\BibitemShut {Stop}%
\bibitem [{\citenamefont {Hay}\ and\ \citenamefont
  {Wadt}(1985{\natexlab{a}})}]{Hay1985a}%
  \BibitemOpen
  \bibfield  {author} {\bibinfo {author} {\bibfnamefont {P.~J.}\ \bibnamefont
  {Hay}}\ and\ \bibinfo {author} {\bibfnamefont {W.~R.}\ \bibnamefont {Wadt}},\
  }\href {\doibase 10.1063/1.448799} {\bibfield  {journal} {\bibinfo  {journal}
  {J. Chem. Phys.}\ }\textbf {\bibinfo {volume} {82}},\ \bibinfo {pages} {270}
  (\bibinfo {year} {1985}{\natexlab{a}})}\BibitemShut {NoStop}%
\bibitem [{\citenamefont {Wadt}\ and\ \citenamefont {Hay}(1985)}]{Wadt1985}%
  \BibitemOpen
  \bibfield  {author} {\bibinfo {author} {\bibfnamefont {W.~R.}\ \bibnamefont
  {Wadt}}\ and\ \bibinfo {author} {\bibfnamefont {P.~J.}\ \bibnamefont {Hay}},\
  }\href {\doibase 10.1063/1.448800} {\bibfield  {journal} {\bibinfo  {journal}
  {J. Chem. Phys.}\ }\textbf {\bibinfo {volume} {82}},\ \bibinfo {pages} {284}
  (\bibinfo {year} {1985})}\BibitemShut {NoStop}%
\bibitem [{\citenamefont {Hay}\ and\ \citenamefont
  {Wadt}(1985{\natexlab{b}})}]{Hay1985b}%
  \BibitemOpen
  \bibfield  {author} {\bibinfo {author} {\bibfnamefont {P.~J.}\ \bibnamefont
  {Hay}}\ and\ \bibinfo {author} {\bibfnamefont {W.~R.}\ \bibnamefont {Wadt}},\
  }\href {\doibase 10.1063/1.448975} {\bibfield  {journal} {\bibinfo  {journal}
  {J. Chem. Phys.}\ }\textbf {\bibinfo {volume} {82}},\ \bibinfo {pages} {299}
  (\bibinfo {year} {1985}{\natexlab{b}})}\BibitemShut {NoStop}%
\bibitem [{\citenamefont {Ditchfield}\ \emph {et~al.}(1971)\citenamefont
  {Ditchfield}, \citenamefont {Hehre},\ and\ \citenamefont {Pople}}]{Pople1}%
  \BibitemOpen
  \bibfield  {author} {\bibinfo {author} {\bibfnamefont {R.}~\bibnamefont
  {Ditchfield}}, \bibinfo {author} {\bibfnamefont {W.~J.}\ \bibnamefont
  {Hehre}}, \ and\ \bibinfo {author} {\bibfnamefont {J.~A.}\ \bibnamefont
  {Pople}},\ }\href {\doibase 10.1063/1.1674902} {\bibfield  {journal}
  {\bibinfo  {journal} {J. Chem. Phys.}\ }\textbf {\bibinfo {volume} {54}},\
  \bibinfo {pages} {724} (\bibinfo {year} {1971})}\BibitemShut {NoStop}%
\bibitem [{\citenamefont {Hehre}\ \emph {et~al.}(1972)\citenamefont {Hehre},
  \citenamefont {Ditchfield},\ and\ \citenamefont {Pople}}]{Pople2}%
  \BibitemOpen
  \bibfield  {author} {\bibinfo {author} {\bibfnamefont {W.~J.}\ \bibnamefont
  {Hehre}}, \bibinfo {author} {\bibfnamefont {R.}~\bibnamefont {Ditchfield}}, \
  and\ \bibinfo {author} {\bibfnamefont {J.~A.}\ \bibnamefont {Pople}},\ }\href
  {\doibase 10.1063/1.1677527} {\bibfield  {journal} {\bibinfo  {journal} {J.
  Chem. Phys.}\ }\textbf {\bibinfo {volume} {56}},\ \bibinfo {pages} {2257}
  (\bibinfo {year} {1972})}\BibitemShut {NoStop}%
\bibitem [{\citenamefont {Hariharan}\ and\ \citenamefont
  {Pople}(1973)}]{Pople3}%
  \BibitemOpen
  \bibfield  {author} {\bibinfo {author} {\bibfnamefont {P.~C.}\ \bibnamefont
  {Hariharan}}\ and\ \bibinfo {author} {\bibfnamefont {J.~A.}\ \bibnamefont
  {Pople}},\ }\href {\doibase 10.1007/BF00533485} {\bibfield  {journal}
  {\bibinfo  {journal} {Theoret. Chim. Acta}\ }\textbf {\bibinfo {volume}
  {28}},\ \bibinfo {pages} {213} (\bibinfo {year} {1973})}\BibitemShut
  {NoStop}%
\bibitem [{\citenamefont {Francl}\ \emph {et~al.}(1982)\citenamefont {Francl},
  \citenamefont {Pietro}, \citenamefont {Hehre}, \citenamefont {Binkley},
  \citenamefont {Gordon}, \citenamefont {DeFrees},\ and\ \citenamefont
  {Pople}}]{Pople4}%
  \BibitemOpen
  \bibfield  {author} {\bibinfo {author} {\bibfnamefont {M.~M.}\ \bibnamefont
  {Francl}}, \bibinfo {author} {\bibfnamefont {W.~J.}\ \bibnamefont {Pietro}},
  \bibinfo {author} {\bibfnamefont {W.~J.}\ \bibnamefont {Hehre}}, \bibinfo
  {author} {\bibfnamefont {J.~S.}\ \bibnamefont {Binkley}}, \bibinfo {author}
  {\bibfnamefont {M.~S.}\ \bibnamefont {Gordon}}, \bibinfo {author}
  {\bibfnamefont {D.~J.}\ \bibnamefont {DeFrees}}, \ and\ \bibinfo {author}
  {\bibfnamefont {J.~A.}\ \bibnamefont {Pople}},\ }\href {\doibase
  10.1063/1.444267} {\bibfield  {journal} {\bibinfo  {journal} {J. Chem.
  Phys.}\ }\textbf {\bibinfo {volume} {77}},\ \bibinfo {pages} {3654} (\bibinfo
  {year} {1982})}\BibitemShut {NoStop}%
\bibitem [{\citenamefont {Bethe}(1931)}]{Bethe1931}%
  \BibitemOpen
  \bibfield  {author} {\bibinfo {author} {\bibfnamefont {H.~Z.}\ \bibnamefont
  {Bethe}},\ }\href {\doibase 10.1007/BF01341708} {\bibfield  {journal}
  {\bibinfo  {journal} {Z. Phys.}\ }\textbf {\bibinfo {volume} {71}},\ \bibinfo
  {pages} {205} (\bibinfo {year} {1931})}\BibitemShut {NoStop}%
\bibitem [{\citenamefont {Yang}\ and\ \citenamefont {Yang}(1966)}]{Yang1966}%
  \BibitemOpen
  \bibfield  {author} {\bibinfo {author} {\bibfnamefont {C.~N.}\ \bibnamefont
  {Yang}}\ and\ \bibinfo {author} {\bibfnamefont {C.~P.}\ \bibnamefont
  {Yang}},\ }\href {\doibase 10.1103/PhysRev.150.321} {\bibfield  {journal}
  {\bibinfo  {journal} {Phys. Rev.}\ }\textbf {\bibinfo {volume} {150}},\
  \bibinfo {pages} {321} (\bibinfo {year} {1966})}\BibitemShut {NoStop}%
\bibitem [{\citenamefont {Schulz}\ and\ \citenamefont
  {Bourbannais}(1983)}]{Schulz1983}%
  \BibitemOpen
  \bibfield  {author} {\bibinfo {author} {\bibfnamefont {H.~J.}\ \bibnamefont
  {Schulz}}\ and\ \bibinfo {author} {\bibfnamefont {C.}~\bibnamefont
  {Bourbannais}},\ }\href {\doibase 10.1103/PhysRevB.27.5856} {\bibfield
  {journal} {\bibinfo  {journal} {Phys. Rev. B}\ }\textbf {\bibinfo {volume}
  {27}},\ \bibinfo {pages} {5856} (\bibinfo {year} {1983})}\BibitemShut
  {NoStop}%
\bibitem [{\citenamefont {Eggert}\ \emph {et~al.}(1994)\citenamefont {Eggert},
  \citenamefont {Affleck},\ and\ \citenamefont {Takahashi}}]{Eggert1994}%
  \BibitemOpen
  \bibfield  {author} {\bibinfo {author} {\bibfnamefont {S.}~\bibnamefont
  {Eggert}}, \bibinfo {author} {\bibfnamefont {I.}~\bibnamefont {Affleck}}, \
  and\ \bibinfo {author} {\bibfnamefont {M.}~\bibnamefont {Takahashi}},\ }\href
  {\doibase 10.1103/PhysRevLett.73.332} {\bibfield  {journal} {\bibinfo
  {journal} {Phys. Rev. Lett.}\ }\textbf {\bibinfo {volume} {73}},\ \bibinfo
  {pages} {332} (\bibinfo {year} {1994})}\BibitemShut {NoStop}%
\bibitem [{\citenamefont {Schulz}(1986)}]{Schulz1986}%
  \BibitemOpen
  \bibfield  {author} {\bibinfo {author} {\bibfnamefont {H.~J.}\ \bibnamefont
  {Schulz}},\ }\href {\doibase 10.1103/PhysRevB.34.6372} {\bibfield  {journal}
  {\bibinfo  {journal} {Phys. Rev. B}\ }\textbf {\bibinfo {volume} {34}},\
  \bibinfo {pages} {6372} (\bibinfo {year} {1986})}\BibitemShut {NoStop}%
\bibitem [{\citenamefont {Affleck}(1998)}]{Affleck1998}%
  \BibitemOpen
  \bibfield  {author} {\bibinfo {author} {\bibfnamefont {I.}~\bibnamefont
  {Affleck}},\ }\href {\doibase 10.1088/0305-4470/31/20/002} {\bibfield
  {journal} {\bibinfo  {journal} {J. Phys. A}\ }\textbf {\bibinfo {volume}
  {31}},\ \bibinfo {pages} {4573} (\bibinfo {year} {1998})}\BibitemShut
  {NoStop}%
\bibitem [{\citenamefont {Barzykin}(2000)}]{Barzykin2000}%
  \BibitemOpen
  \bibfield  {author} {\bibinfo {author} {\bibfnamefont {V.}~\bibnamefont
  {Barzykin}},\ }\href {\doibase 10.1088/0953-8984/12/9/309} {\bibfield
  {journal} {\bibinfo  {journal} {J. Phys.: Condens. Matter}\ }\textbf
  {\bibinfo {volume} {12}},\ \bibinfo {pages} {2053} (\bibinfo {year}
  {2000})}\BibitemShut {NoStop}%
\bibitem [{\citenamefont {Lukyanov}(1998)}]{Lukyanov1998}%
  \BibitemOpen
  \bibfield  {author} {\bibinfo {author} {\bibfnamefont {S.}~\bibnamefont
  {Lukyanov}},\ }\href {\doibase 10.1016/S0550-3213(98)00249-1} {\bibfield
  {journal} {\bibinfo  {journal} {Nucl. Phys. B}\ }\textbf {\bibinfo {volume}
  {522}},\ \bibinfo {pages} {533} (\bibinfo {year} {1998})}\BibitemShut
  {NoStop}%
\bibitem [{\citenamefont {Luther}\ and\ \citenamefont
  {Peschel}(1974)}]{Luther1974}%
  \BibitemOpen
  \bibfield  {author} {\bibinfo {author} {\bibfnamefont {A.}~\bibnamefont
  {Luther}}\ and\ \bibinfo {author} {\bibfnamefont {I.}~\bibnamefont
  {Peschel}},\ }\href {\doibase 10.1103/PhysRevB.9.2911} {\bibfield  {journal}
  {\bibinfo  {journal} {Phys. Rev. B}\ }\textbf {\bibinfo {volume} {9}},\
  \bibinfo {pages} {2911} (\bibinfo {year} {1974})}\BibitemShut {NoStop}%
\bibitem [{\citenamefont {Tsvelik}(2003)}]{TsvelikBook}%
  \BibitemOpen
  \bibfield  {author} {\bibinfo {author} {\bibfnamefont {A.~M.}\ \bibnamefont
  {Tsvelik}},\ }\href@noop {} {\emph {\bibinfo {title} {Quantum Field Theory in
  Condensed Matter Physics}}},\ Vol.~\bibinfo {volume} {2}\ (\bibinfo
  {publisher} {Cambridge University Press},\ \bibinfo {address} {Cambridge,
  UK},\ \bibinfo {year} {2003})\BibitemShut {NoStop}%
\bibitem [{\citenamefont {Scalapino}\ \emph {et~al.}(1975)\citenamefont
  {Scalapino}, \citenamefont {Imry},\ and\ \citenamefont
  {Pincus}}]{Scalapino1975}%
  \BibitemOpen
  \bibfield  {author} {\bibinfo {author} {\bibfnamefont {D.~J.}\ \bibnamefont
  {Scalapino}}, \bibinfo {author} {\bibfnamefont {Y.}~\bibnamefont {Imry}}, \
  and\ \bibinfo {author} {\bibfnamefont {P.}~\bibnamefont {Pincus}},\ }\href
  {\doibase 10.1103/PhysRevB.11.2042} {\bibfield  {journal} {\bibinfo
  {journal} {Phys. Rev. B}\ }\textbf {\bibinfo {volume} {11}},\ \bibinfo
  {pages} {2042} (\bibinfo {year} {1975})}\BibitemShut {NoStop}%
\bibitem [{\citenamefont {Schulz}(1996)}]{Schulz1996}%
  \BibitemOpen
  \bibfield  {author} {\bibinfo {author} {\bibfnamefont {H.~J.}\ \bibnamefont
  {Schulz}},\ }\href {\doibase 10.1103/PhysRevLett.77.2790} {\bibfield
  {journal} {\bibinfo  {journal} {Phys. Rev. Lett.}\ }\textbf {\bibinfo
  {volume} {77}},\ \bibinfo {pages} {2790} (\bibinfo {year}
  {1996})}\BibitemShut {NoStop}%
\bibitem [{\citenamefont {Essler}\ \emph {et~al.}(1997)\citenamefont {Essler},
  \citenamefont {Tsvelik},\ and\ \citenamefont {Delfino}}]{Essler1997}%
  \BibitemOpen
  \bibfield  {author} {\bibinfo {author} {\bibfnamefont {F.~H.~L.}\
  \bibnamefont {Essler}}, \bibinfo {author} {\bibfnamefont {A.~M.}\
  \bibnamefont {Tsvelik}}, \ and\ \bibinfo {author} {\bibfnamefont
  {G.}~\bibnamefont {Delfino}},\ }\href {\doibase 10.1103/PhysRevB.56.11001}
  {\bibfield  {journal} {\bibinfo  {journal} {Phys. Rev. B}\ }\textbf {\bibinfo
  {volume} {56}},\ \bibinfo {pages} {11001} (\bibinfo {year}
  {1997})}\BibitemShut {NoStop}%
\bibitem [{\citenamefont {Yasuda}\ \emph {et~al.}(2005)\citenamefont {Yasuda},
  \citenamefont {Todo}, \citenamefont {Hukushima}, \citenamefont {Alet},
  \citenamefont {Keller}, \citenamefont {Troyer},\ and\ \citenamefont
  {Takayama}}]{Yasuda2005}%
  \BibitemOpen
  \bibfield  {author} {\bibinfo {author} {\bibfnamefont {C.}~\bibnamefont
  {Yasuda}}, \bibinfo {author} {\bibfnamefont {S.}~\bibnamefont {Todo}},
  \bibinfo {author} {\bibfnamefont {K.}~\bibnamefont {Hukushima}}, \bibinfo
  {author} {\bibfnamefont {F.}~\bibnamefont {Alet}}, \bibinfo {author}
  {\bibfnamefont {M.}~\bibnamefont {Keller}}, \bibinfo {author} {\bibfnamefont
  {M.}~\bibnamefont {Troyer}}, \ and\ \bibinfo {author} {\bibfnamefont
  {H.}~\bibnamefont {Takayama}},\ }\href {\doibase
  10.1103/PhysRevLett.94.217201} {\bibfield  {journal} {\bibinfo  {journal}
  {Phys. Rev. Lett.}\ }\textbf {\bibinfo {volume} {94}},\ \bibinfo {pages}
  {217201} (\bibinfo {year} {2005})}\BibitemShut {NoStop}%
\bibitem [{\citenamefont {Starykh}\ \emph {et~al.}(2010)\citenamefont
  {Starykh}, \citenamefont {Katsura},\ and\ \citenamefont
  {Balents}}]{Starykh2010}%
  \BibitemOpen
  \bibfield  {author} {\bibinfo {author} {\bibfnamefont {O.~A.}\ \bibnamefont
  {Starykh}}, \bibinfo {author} {\bibfnamefont {H.}~\bibnamefont {Katsura}}, \
  and\ \bibinfo {author} {\bibfnamefont {L.}~\bibnamefont {Balents}},\ }\href
  {\doibase 10.1103/PhysRevB.82.014421} {\bibfield  {journal} {\bibinfo
  {journal} {Phys. Rev. B}\ }\textbf {\bibinfo {volume} {82}},\ \bibinfo
  {pages} {014421} (\bibinfo {year} {2010})}\BibitemShut {NoStop}%
\bibitem [{\citenamefont {Bonner}\ and\ \citenamefont
  {Fisher}(1964)}]{Bonner1964}%
  \BibitemOpen
  \bibfield  {author} {\bibinfo {author} {\bibfnamefont {J.~C.}\ \bibnamefont
  {Bonner}}\ and\ \bibinfo {author} {\bibfnamefont {M.~E.}\ \bibnamefont
  {Fisher}},\ }\href {\doibase 10.1103/PhysRev.135.A640} {\bibfield  {journal}
  {\bibinfo  {journal} {Phys. Rev.}\ }\textbf {\bibinfo {volume} {135}},\
  \bibinfo {pages} {A640} (\bibinfo {year} {1964})}\BibitemShut {NoStop}%
\bibitem [{\citenamefont {Estes}\ \emph {et~al.}(1977)\citenamefont {Estes},
  \citenamefont {Gavel}, \citenamefont {Hatfield},\ and\ \citenamefont
  {Hodgson}}]{Estes1977}%
  \BibitemOpen
  \bibfield  {author} {\bibinfo {author} {\bibfnamefont {W.~E.}\ \bibnamefont
  {Estes}}, \bibinfo {author} {\bibfnamefont {D.~P.}\ \bibnamefont {Gavel}},
  \bibinfo {author} {\bibfnamefont {W.~E.}\ \bibnamefont {Hatfield}}, \ and\
  \bibinfo {author} {\bibfnamefont {D.~J.}\ \bibnamefont {Hodgson}},\ }\href
  {\doibase 10.1021/ic50184a005} {\bibfield  {journal} {\bibinfo  {journal}
  {Inorg. Chem.}\ }\textbf {\bibinfo {volume} {17}},\ \bibinfo {pages} {1415}
  (\bibinfo {year} {1977})}\BibitemShut {NoStop}%
\bibitem [{\citenamefont {Ami}\ \emph {et~al.}(1995)\citenamefont {Ami},
  \citenamefont {Crawford}, \citenamefont {Harlow}, \citenamefont {Wang},
  \citenamefont {Johnston}, \citenamefont {Huang},\ and\ \citenamefont
  {Erwin}}]{Ami1995}%
  \BibitemOpen
  \bibfield  {author} {\bibinfo {author} {\bibfnamefont {T.}~\bibnamefont
  {Ami}}, \bibinfo {author} {\bibfnamefont {M.~K.}\ \bibnamefont {Crawford}},
  \bibinfo {author} {\bibfnamefont {R.~L.}\ \bibnamefont {Harlow}}, \bibinfo
  {author} {\bibfnamefont {Z.~R.}\ \bibnamefont {Wang}}, \bibinfo {author}
  {\bibfnamefont {D.~C.}\ \bibnamefont {Johnston}}, \bibinfo {author}
  {\bibfnamefont {Q.}~\bibnamefont {Huang}}, \ and\ \bibinfo {author}
  {\bibfnamefont {R.~W.}\ \bibnamefont {Erwin}},\ }\href {\doibase
  10.1103/PhysRevB.51.5994} {\bibfield  {journal} {\bibinfo  {journal} {Phys.
  Rev. B}\ }\textbf {\bibinfo {volume} {51}},\ \bibinfo {pages} {5994}
  (\bibinfo {year} {1995})}\BibitemShut {NoStop}%
\bibitem [{\citenamefont {Moreno}\ \emph {et~al.}(2013)\citenamefont {Moreno},
  \citenamefont {Arrue}, \citenamefont {Saavedra}, \citenamefont {Pivan},
  \citenamefont {Pe\~aa},\ and\ \citenamefont {Roisnell}}]{Moreno2013}%
  \BibitemOpen
  \bibfield  {author} {\bibinfo {author} {\bibfnamefont {Y.}~\bibnamefont
  {Moreno}}, \bibinfo {author} {\bibfnamefont {R.}~\bibnamefont {Arrue}},
  \bibinfo {author} {\bibfnamefont {R.}~\bibnamefont {Saavedra}}, \bibinfo
  {author} {\bibfnamefont {J.-Y.}\ \bibnamefont {Pivan}}, \bibinfo {author}
  {\bibfnamefont {O.}~\bibnamefont {Pe\~aa}}, \ and\ \bibinfo {author}
  {\bibfnamefont {T.}~\bibnamefont {Roisnell}},\ }\href@noop {} {\bibfield
  {journal} {\bibinfo  {journal} {{J. Chil. Chem. Soc.}}\ }\textbf {\bibinfo
  {volume} {58}},\ \bibinfo {pages} {2122 } (\bibinfo {year}
  {2013})}\BibitemShut {NoStop}%
\bibitem [{Egg()}]{EggertDat}%
  \BibitemOpen
  \href@noop {} {}\bibinfo {note} {See
  https://www.physik.uni-kl.de/eggert/papers/susceptibility.dat for the
  numerical data in Ref. \citenum{Eggert1994}}\BibitemShut {NoStop}%
\bibitem [{\citenamefont {Haldane}(1980)}]{Haldane1980}%
  \BibitemOpen
  \bibfield  {author} {\bibinfo {author} {\bibfnamefont {F.~D.~M.}\
  \bibnamefont {Haldane}},\ }\href {\doibase 10.1103/PhysRevLett.45.1358}
  {\bibfield  {journal} {\bibinfo  {journal} {Phys. Rev. Lett.}\ }\textbf
  {\bibinfo {volume} {45}},\ \bibinfo {pages} {1358} (\bibinfo {year}
  {1980})}\BibitemShut {NoStop}%
\bibitem [{\citenamefont {Barzykin}(2001)}]{Barzykin2001}%
  \BibitemOpen
  \bibfield  {author} {\bibinfo {author} {\bibfnamefont {V.}~\bibnamefont
  {Barzykin}},\ }\href {\doibase 10.1103/PhysRevB.63.140412} {\bibfield
  {journal} {\bibinfo  {journal} {Phys. Rev. B}\ }\textbf {\bibinfo {volume}
  {63}},\ \bibinfo {pages} {140412} (\bibinfo {year} {2001})}\BibitemShut
  {NoStop}%
\end{thebibliography}%


\providecommand*{\mcitethebibliography}{\thebibliography}
\csname @ifundefined\endcsname{endmcitethebibliography}
{\let\endmcitethebibliography\endthebibliography}{}
\begin{mcitethebibliography}{47}
\providecommand*{\natexlab}[1]{#1}
\providecommand*{\mciteSetBstSublistMode}[1]{}
\providecommand*{\mciteSetBstMaxWidthForm}[2]{}
\providecommand*{\mciteBstWouldAddEndPuncttrue}
  {\def\EndOfBibitem{\unskip.}}
\providecommand*{\mciteBstWouldAddEndPunctfalse}
  {\let\EndOfBibitem\relax}
\providecommand*{\mciteSetBstMidEndSepPunct}[3]{}
\providecommand*{\mciteSetBstSublistLabelBeginEnd}[3]{}
\providecommand*{\EndOfBibitem}{}
\mciteSetBstSublistMode{f}
\mciteSetBstMaxWidthForm{subitem}
{\alph{mcitesubitemcount})}
\mciteSetBstSublistLabelBeginEnd{\mcitemaxwidthsubitemform\space}
{\relax}{\relax}

\bibitem[Bethe(1931)]{Bethe1931}
H.~Z. Bethe, \emph{Z. Phys.} \textbf{1931}, \emph{71}, 205--226\relax
\mciteBstWouldAddEndPuncttrue
\mciteSetBstMidEndSepPunct{\mcitedefaultmidpunct}
{\mcitedefaultendpunct}{\mcitedefaultseppunct}\relax
\EndOfBibitem
\bibitem[Yang and Yang(1966)]{Yang1966}
C.~N. Yang, C.~P. Yang, \emph{Phys. Rev.} \textbf{1966}, \emph{150},
  321--327\relax
\mciteBstWouldAddEndPuncttrue
\mciteSetBstMidEndSepPunct{\mcitedefaultmidpunct}
{\mcitedefaultendpunct}{\mcitedefaultseppunct}\relax
\EndOfBibitem
\bibitem[Schulz and Bourbannais(1983)]{Schulz1983}
H.~J. Schulz, C.~Bourbannais, \emph{Phys. Rev. B} \textbf{1983}, \emph{27},
  5856--5859\relax
\mciteBstWouldAddEndPuncttrue
\mciteSetBstMidEndSepPunct{\mcitedefaultmidpunct}
{\mcitedefaultendpunct}{\mcitedefaultseppunct}\relax
\EndOfBibitem
\bibitem[Eggert \emph{et~al.}(1994)Eggert, Affleck, and Takahashi]{Eggert1994}
S.~Eggert, I.~Affleck, M.~Takahashi, \emph{Phys. Rev. Lett.} \textbf{1994},
  \emph{73}, 332--335\relax
\mciteBstWouldAddEndPuncttrue
\mciteSetBstMidEndSepPunct{\mcitedefaultmidpunct}
{\mcitedefaultendpunct}{\mcitedefaultseppunct}\relax
\EndOfBibitem
\bibitem[Schulz(1986)]{Schulz1986}
H.~J. Schulz, \emph{Phys. Rev. B} \textbf{1986}, \emph{34}, 6372--6385\relax
\mciteBstWouldAddEndPuncttrue
\mciteSetBstMidEndSepPunct{\mcitedefaultmidpunct}
{\mcitedefaultendpunct}{\mcitedefaultseppunct}\relax
\EndOfBibitem
\bibitem[Affleck(1998)]{Affleck1998}
I.~Affleck, \emph{J. Phys. A} \textbf{1998}, \emph{31}, 4573\relax
\mciteBstWouldAddEndPuncttrue
\mciteSetBstMidEndSepPunct{\mcitedefaultmidpunct}
{\mcitedefaultendpunct}{\mcitedefaultseppunct}\relax
\EndOfBibitem
\bibitem[Barzykin(2000)]{Barzykin2000}
V.~Barzykin, \emph{J. Phys.: Condens. Matter} \textbf{2000}, \emph{12},
  2053\relax
\mciteBstWouldAddEndPuncttrue
\mciteSetBstMidEndSepPunct{\mcitedefaultmidpunct}
{\mcitedefaultendpunct}{\mcitedefaultseppunct}\relax
\EndOfBibitem
\bibitem[Lukyanov(1998)]{Lukyanov1998}
S.~Lukyanov, \emph{Nucl. Phys. B} \textbf{1998}, \emph{522}, 533--549\relax
\mciteBstWouldAddEndPuncttrue
\mciteSetBstMidEndSepPunct{\mcitedefaultmidpunct}
{\mcitedefaultendpunct}{\mcitedefaultseppunct}\relax
\EndOfBibitem
\bibitem[Luther and Peschel(1974)]{Luther1974}
A.~Luther, I.~Peschel, \emph{Phys. Rev. B} \textbf{1974}, \emph{9},
  2911--2919\relax
\mciteBstWouldAddEndPuncttrue
\mciteSetBstMidEndSepPunct{\mcitedefaultmidpunct}
{\mcitedefaultendpunct}{\mcitedefaultseppunct}\relax
\EndOfBibitem
\bibitem[Tsvelik(2003)]{TsvelikBook}
A.~M. Tsvelik, \emph{Quantum Field Theory in Condensed Matter Physics},
  \emph{Vol.~2}, Cambridge University Press, Cambridge, UK, \textbf{2003}\relax
\mciteBstWouldAddEndPuncttrue
\mciteSetBstMidEndSepPunct{\mcitedefaultmidpunct}
{\mcitedefaultendpunct}{\mcitedefaultseppunct}\relax
\EndOfBibitem
\bibitem[Barzykin(2001)]{Barzykin2001}
V.~Barzykin, \emph{Phys. Rev. B} \textbf{2001}, \emph{63}, 140412\relax
\mciteBstWouldAddEndPuncttrue
\mciteSetBstMidEndSepPunct{\mcitedefaultmidpunct}
{\mcitedefaultendpunct}{\mcitedefaultseppunct}\relax
\EndOfBibitem
\bibitem[Scalapino \emph{et~al.}(1975)Scalapino, Imry, and
  Pincus]{Scalapino1975}
D.~J. Scalapino, Y.~Imry, P.~Pincus, \emph{Phys. Rev. B} \textbf{1975},
  \emph{11}, 2042--2048\relax
\mciteBstWouldAddEndPuncttrue
\mciteSetBstMidEndSepPunct{\mcitedefaultmidpunct}
{\mcitedefaultendpunct}{\mcitedefaultseppunct}\relax
\EndOfBibitem
\bibitem[Schulz(1996)]{Schulz1996}
H.~J. Schulz, \emph{Phys. Rev. Lett.} \textbf{1996}, \emph{77},
  2790--2793\relax
\mciteBstWouldAddEndPuncttrue
\mciteSetBstMidEndSepPunct{\mcitedefaultmidpunct}
{\mcitedefaultendpunct}{\mcitedefaultseppunct}\relax
\EndOfBibitem
\bibitem[Essler \emph{et~al.}(1997)Essler, Tsvelik, and Delfino]{Essler1997}
F.~H.~L. Essler, A.~M. Tsvelik, G.~Delfino, \emph{Phys. Rev. B} \textbf{1997},
  \emph{56}, 11001\relax
\mciteBstWouldAddEndPuncttrue
\mciteSetBstMidEndSepPunct{\mcitedefaultmidpunct}
{\mcitedefaultendpunct}{\mcitedefaultseppunct}\relax
\EndOfBibitem
\bibitem[Bocquet \emph{et~al.}(2001)Bocquet, Essler, Tsvelik, and
  Gogolin]{Bocquet2001}
M.~Bocquet, F.~H.~L. Essler, A.~M. Tsvelik, A.~O. Gogolin, \emph{Phys. Rev. B}
  \textbf{2001}, \emph{64}, 094425\relax
\mciteBstWouldAddEndPuncttrue
\mciteSetBstMidEndSepPunct{\mcitedefaultmidpunct}
{\mcitedefaultendpunct}{\mcitedefaultseppunct}\relax
\EndOfBibitem
\bibitem[Worthy \emph{et~al.}(2018)Worthy, Grosjean, Pfrunder, Xu, Yan,
  Edwards, Clegg, and McMurtrie]{Worthy2017}
A.~Worthy, A.~Grosjean, M.~C. Pfrunder, Y.~Xu, C.~Yan, G.~Edwards, J.~K. Clegg,
  J.~C. McMurtrie, \emph{Nat. Chem.} \textbf{2018}, \emph{10}, 65--69\relax
\mciteBstWouldAddEndPuncttrue
\mciteSetBstMidEndSepPunct{\mcitedefaultmidpunct}
{\mcitedefaultendpunct}{\mcitedefaultseppunct}\relax
\EndOfBibitem
\bibitem[Egg()]{EggertDat}
See https://www.physik.uni-kl.de/eggert/papers/susceptibility.dat for the
  numerical data in Ref. \citenum{Eggert1994}\relax
\mciteBstWouldAddEndPuncttrue
\mciteSetBstMidEndSepPunct{\mcitedefaultmidpunct}
{\mcitedefaultendpunct}{\mcitedefaultseppunct}\relax
\EndOfBibitem
\bibitem[Bonner and Fisher(1964)]{Bonner1964}
J.~C. Bonner, M.~E. Fisher, \emph{Phys. Rev.} \textbf{1964}, \emph{135},
  A640--A658\relax
\mciteBstWouldAddEndPuncttrue
\mciteSetBstMidEndSepPunct{\mcitedefaultmidpunct}
{\mcitedefaultendpunct}{\mcitedefaultseppunct}\relax
\EndOfBibitem
\bibitem[Estes \emph{et~al.}(1977)Estes, Gavel, Hatfield, and
  Hodgson]{Estes1977}
W.~E. Estes, D.~P. Gavel, W.~E. Hatfield, D.~J. Hodgson, \emph{Inorg. Chem.}
  \textbf{1977}, \emph{17}, 1415--1421\relax
\mciteBstWouldAddEndPuncttrue
\mciteSetBstMidEndSepPunct{\mcitedefaultmidpunct}
{\mcitedefaultendpunct}{\mcitedefaultseppunct}\relax
\EndOfBibitem
\bibitem[Ami \emph{et~al.}(1995)Ami, Crawford, Harlow, Wang, Johnston, Huang,
  and Erwin]{Ami1995}
T.~Ami, M.~K. Crawford, R.~L. Harlow, Z.~R. Wang, D.~C. Johnston, Q.~Huang,
  R.~W. Erwin, \emph{Phys. Rev. B} \textbf{1995}, \emph{51}, 5994--6001\relax
\mciteBstWouldAddEndPuncttrue
\mciteSetBstMidEndSepPunct{\mcitedefaultmidpunct}
{\mcitedefaultendpunct}{\mcitedefaultseppunct}\relax
\EndOfBibitem
\bibitem[Moreno \emph{et~al.}(2013)Moreno, Arrue, Saavedra, Pivan, Pe\~aa, and
  Roisnell]{Moreno2013}
Y.~Moreno, R.~Arrue, R.~Saavedra, J.-Y. Pivan, O.~Pe\~aa, T.~Roisnell,
  \emph{{J. Chil. Chem. Soc.}} \textbf{2013}, \emph{58}, 2122 -- 2124\relax
\mciteBstWouldAddEndPuncttrue
\mciteSetBstMidEndSepPunct{\mcitedefaultmidpunct}
{\mcitedefaultendpunct}{\mcitedefaultseppunct}\relax
\EndOfBibitem
\bibitem[Lovesey(1986)]{Lovesey}
S.~W. Lovesey, \emph{Theory of neutron scattering from condensed matter},
  Oxford University Press, Oxford, \textbf{1986}\relax
\mciteBstWouldAddEndPuncttrue
\mciteSetBstMidEndSepPunct{\mcitedefaultmidpunct}
{\mcitedefaultendpunct}{\mcitedefaultseppunct}\relax
\EndOfBibitem
\bibitem[Frisch \emph{et~al.}()Frisch\emph{et~al.}]{g09}
M.~J. Frisch \emph{et~al.}, \emph{Gaussian~09 {R}evision {E}.01}, Gaussian Inc.
  Wallingford CT 2009\relax
\mciteBstWouldAddEndPuncttrue
\mciteSetBstMidEndSepPunct{\mcitedefaultmidpunct}
{\mcitedefaultendpunct}{\mcitedefaultseppunct}\relax
\EndOfBibitem
\bibitem[Becke(1993)]{B3LYPa}
A.~D. Becke, \emph{J. Chem. Phys.} \textbf{1993}, \emph{98}, 5648--5652\relax
\mciteBstWouldAddEndPuncttrue
\mciteSetBstMidEndSepPunct{\mcitedefaultmidpunct}
{\mcitedefaultendpunct}{\mcitedefaultseppunct}\relax
\EndOfBibitem
\bibitem[Stephens \emph{et~al.}(1994)Stephens, Devlin, Chabalowski, and
  Frisch]{B3LYPb}
P.~J. Stephens, F.~J. Devlin, C.~F. Chabalowski, M.~J. Frisch, \emph{J. Phys.
  Chem.} \textbf{1994}, \emph{98}, 11623--11627\relax
\mciteBstWouldAddEndPuncttrue
\mciteSetBstMidEndSepPunct{\mcitedefaultmidpunct}
{\mcitedefaultendpunct}{\mcitedefaultseppunct}\relax
\EndOfBibitem
\bibitem[Ditchfield \emph{et~al.}(1971)Ditchfield, Hehre, and Pople]{Pople1}
R.~Ditchfield, W.~J. Hehre, J.~A. Pople, \emph{J. Chem. Phys.} \textbf{1971},
  \emph{54}, 724--728\relax
\mciteBstWouldAddEndPuncttrue
\mciteSetBstMidEndSepPunct{\mcitedefaultmidpunct}
{\mcitedefaultendpunct}{\mcitedefaultseppunct}\relax
\EndOfBibitem
\bibitem[Hehre \emph{et~al.}(1972)Hehre, Ditchfield, and Pople]{Pople2}
W.~J. Hehre, R.~Ditchfield, J.~A. Pople, \emph{J. Chem. Phys.} \textbf{1972},
  \emph{56}, 2257--2261\relax
\mciteBstWouldAddEndPuncttrue
\mciteSetBstMidEndSepPunct{\mcitedefaultmidpunct}
{\mcitedefaultendpunct}{\mcitedefaultseppunct}\relax
\EndOfBibitem
\bibitem[Hariharan and Pople(1973)]{Pople3}
P.~C. Hariharan, J.~A. Pople, \emph{Theoret. Chim. Acta} \textbf{1973},
  \emph{28}, 213--222\relax
\mciteBstWouldAddEndPuncttrue
\mciteSetBstMidEndSepPunct{\mcitedefaultmidpunct}
{\mcitedefaultendpunct}{\mcitedefaultseppunct}\relax
\EndOfBibitem
\bibitem[Francl \emph{et~al.}(1982)Francl, Pietro, Hehre, Binkley, Gordon,
  DeFrees, and Pople]{Pople4}
M.~M. Francl, W.~J. Pietro, W.~J. Hehre, J.~S. Binkley, M.~S. Gordon, D.~J.
  DeFrees, J.~A. Pople, \emph{J. Chem. Phys.} \textbf{1982}, \emph{77},
  3654--3665\relax
\mciteBstWouldAddEndPuncttrue
\mciteSetBstMidEndSepPunct{\mcitedefaultmidpunct}
{\mcitedefaultendpunct}{\mcitedefaultseppunct}\relax
\EndOfBibitem
\bibitem[Rassolov \emph{et~al.}(1998)Rassolov, Pople, Ratner, and
  Windus]{Pople_3rdRow}
V.~A. Rassolov, J.~A. Pople, M.~A. Ratner, T.~L. Windus, \emph{J. Chem. Phys.}
  \textbf{1998}, \emph{109}, 1223--1229\relax
\mciteBstWouldAddEndPuncttrue
\mciteSetBstMidEndSepPunct{\mcitedefaultmidpunct}
{\mcitedefaultendpunct}{\mcitedefaultseppunct}\relax
\EndOfBibitem
\bibitem[Krishnan \emph{et~al.}(1980)Krishnan, Binkley, Seeger, and
  Pople]{6311Ga}
R.~Krishnan, J.~S. Binkley, R.~Seeger, J.~A. Pople, \emph{J. Chem. Phys.}
  \textbf{1980}, \emph{72}, 650--654\relax
\mciteBstWouldAddEndPuncttrue
\mciteSetBstMidEndSepPunct{\mcitedefaultmidpunct}
{\mcitedefaultendpunct}{\mcitedefaultseppunct}\relax
\EndOfBibitem
\bibitem[Wachters(1970)]{6311Gb}
A.~J.~H. Wachters, \emph{J. Chem. Phys.} \textbf{1970}, \emph{52},
  1033--1036\relax
\mciteBstWouldAddEndPuncttrue
\mciteSetBstMidEndSepPunct{\mcitedefaultmidpunct}
{\mcitedefaultendpunct}{\mcitedefaultseppunct}\relax
\EndOfBibitem
\bibitem[Hay(1977)]{6311Gc}
P.~J. Hay, \emph{J. Chem. Phys.} \textbf{1977}, \emph{66}, 4377--4384\relax
\mciteBstWouldAddEndPuncttrue
\mciteSetBstMidEndSepPunct{\mcitedefaultmidpunct}
{\mcitedefaultendpunct}{\mcitedefaultseppunct}\relax
\EndOfBibitem
\bibitem[Raghavachari and Trucks(1989)]{6311Gd}
K.~Raghavachari, G.~W. Trucks, \emph{J. Chem. Phys.} \textbf{1989}, \emph{91},
  1062--1065\relax
\mciteBstWouldAddEndPuncttrue
\mciteSetBstMidEndSepPunct{\mcitedefaultmidpunct}
{\mcitedefaultendpunct}{\mcitedefaultseppunct}\relax
\EndOfBibitem
\bibitem[Sch\"{a}fer \emph{et~al.}(1992)Sch\"{a}fer, Horn, and Ahlrichs]{TZVPa}
A.~Sch\"{a}fer, H.~Horn, R.~Ahlrichs, \emph{J. Chem. Phys.} \textbf{1992},
  \emph{97}, 2571--2577\relax
\mciteBstWouldAddEndPuncttrue
\mciteSetBstMidEndSepPunct{\mcitedefaultmidpunct}
{\mcitedefaultendpunct}{\mcitedefaultseppunct}\relax
\EndOfBibitem
\bibitem[Sch\"{a}fer \emph{et~al.}(1994)Sch\"{a}fer, Huber, and
  Ahlrichs]{TZVPb}
A.~Sch\"{a}fer, C.~Huber, R.~Ahlrichs, \emph{J. Chem. Phys.} \textbf{1994},
  \emph{100}, 5829--5835\relax
\mciteBstWouldAddEndPuncttrue
\mciteSetBstMidEndSepPunct{\mcitedefaultmidpunct}
{\mcitedefaultendpunct}{\mcitedefaultseppunct}\relax
\EndOfBibitem
\bibitem[{Dunning Jr.} and Hay(1977)]{Dunning1977}
T.~H. {Dunning Jr.}, P.~J. Hay in \emph{Methods of Electronic Structure
  Theory}, \emph{Vol.~2}, Plenum Press, 3rd ed., \textbf{1977}\relax
\mciteBstWouldAddEndPuncttrue
\mciteSetBstMidEndSepPunct{\mcitedefaultmidpunct}
{\mcitedefaultendpunct}{\mcitedefaultseppunct}\relax
\EndOfBibitem
\bibitem[Hay and Wadt(1985)]{Hay1985a}
P.~J. Hay, W.~R. Wadt, \emph{J. Chem. Phys.} \textbf{1985}, \emph{82},
  270--283\relax
\mciteBstWouldAddEndPuncttrue
\mciteSetBstMidEndSepPunct{\mcitedefaultmidpunct}
{\mcitedefaultendpunct}{\mcitedefaultseppunct}\relax
\EndOfBibitem
\bibitem[Wadt and Hay(1985)]{Wadt1985}
W.~R. Wadt, P.~J. Hay, \emph{J. Chem. Phys.} \textbf{1985}, \emph{82},
  284--298\relax
\mciteBstWouldAddEndPuncttrue
\mciteSetBstMidEndSepPunct{\mcitedefaultmidpunct}
{\mcitedefaultendpunct}{\mcitedefaultseppunct}\relax
\EndOfBibitem
\bibitem[Hay and Wadt(1985)]{Hay1985b}
P.~J. Hay, W.~R. Wadt, \emph{J. Chem. Phys.} \textbf{1985}, \emph{82},
  299--310\relax
\mciteBstWouldAddEndPuncttrue
\mciteSetBstMidEndSepPunct{\mcitedefaultmidpunct}
{\mcitedefaultendpunct}{\mcitedefaultseppunct}\relax
\EndOfBibitem
\bibitem[Dunning(1989)]{Dunning1}
T.~H. Dunning, \emph{J. Chem. Phys.} \textbf{1989}, \emph{90}, 1007--1023\relax
\mciteBstWouldAddEndPuncttrue
\mciteSetBstMidEndSepPunct{\mcitedefaultmidpunct}
{\mcitedefaultendpunct}{\mcitedefaultseppunct}\relax
\EndOfBibitem
\bibitem[Kendall \emph{et~al.}(1992)Kendall, Dunning, and Harrison]{Dunning2}
R.~A. Kendall, T.~H. Dunning, R.~J. Harrison, \emph{J. Chem. Phys.}
  \textbf{1992}, \emph{96}, 6796--6806\relax
\mciteBstWouldAddEndPuncttrue
\mciteSetBstMidEndSepPunct{\mcitedefaultmidpunct}
{\mcitedefaultendpunct}{\mcitedefaultseppunct}\relax
\EndOfBibitem
\bibitem[Peterson \emph{et~al.}(1994)Peterson, Woon, and Dunning]{Dunning3}
K.~A. Peterson, D.~E. Woon, T.~H. Dunning, \emph{J. Chem. Phys.} \textbf{1994},
  \emph{100}, 7410--7415\relax
\mciteBstWouldAddEndPuncttrue
\mciteSetBstMidEndSepPunct{\mcitedefaultmidpunct}
{\mcitedefaultendpunct}{\mcitedefaultseppunct}\relax
\EndOfBibitem
\bibitem[Perdew \emph{et~al.}(1996)Perdew, Burke, and Ernzerhof]{PBEa}
J.~P. Perdew, K.~Burke, M.~Ernzerhof, \emph{Phys. Rev. Lett.} \textbf{1996},
  \emph{77}, 3865--3868\relax
\mciteBstWouldAddEndPuncttrue
\mciteSetBstMidEndSepPunct{\mcitedefaultmidpunct}
{\mcitedefaultendpunct}{\mcitedefaultseppunct}\relax
\EndOfBibitem
\bibitem[Perdew \emph{et~al.}(1997)Perdew, Burke, and Ernzerhof]{PBEb}
J.~P. Perdew, K.~Burke, M.~Ernzerhof, \emph{Phys. Rev. Lett.} \textbf{1997},
  \emph{78}, 1396--1396\relax
\mciteBstWouldAddEndPuncttrue
\mciteSetBstMidEndSepPunct{\mcitedefaultmidpunct}
{\mcitedefaultendpunct}{\mcitedefaultseppunct}\relax
\EndOfBibitem
\bibitem[Tao \emph{et~al.}(2003)Tao, Perdew, Staroverov, and Scuseria]{TPSSha}
J.~Tao, J.~P. Perdew, V.~N. Staroverov, G.~E. Scuseria, \emph{Phys. Rev. Lett.}
  \textbf{2003}, \emph{91}, 146401\relax
\mciteBstWouldAddEndPuncttrue
\mciteSetBstMidEndSepPunct{\mcitedefaultmidpunct}
{\mcitedefaultendpunct}{\mcitedefaultseppunct}\relax
\EndOfBibitem
\bibitem[Staroverov \emph{et~al.}(2004)Staroverov, Scuseria, Tao, and
  Perdew]{TPSShb}
V.~N. Staroverov, G.~E. Scuseria, J.~Tao, J.~P. Perdew, \emph{J. Chem. Phys.}
  \textbf{2004}, \emph{121}, 11507--11507\relax
\mciteBstWouldAddEndPuncttrue
\mciteSetBstMidEndSepPunct{\mcitedefaultmidpunct}
{\mcitedefaultendpunct}{\mcitedefaultseppunct}\relax
\EndOfBibitem
\end{mcitethebibliography}

\end{document}


\title[]
{-- Supplementary Information -- \\ Towards mechanomagnetics in elastic crystals: insights from [Cu(acac)$_2$]}
\author{E. P. Kenny}
\affiliation{School of Mathematics and Physics, The University of Queensland, Brisbane, Queensland, Australia}
\email{elisekenny@gmail.com}
\author{A. C. Jacko}
\affiliation{School of Mathematics and Physics, The University of Queensland, Brisbane, Queensland, Australia}
\author{B. J. Powell}
\affiliation{School of Mathematics and Physics, The University of Queensland, Brisbane, Queensland, Australia}

\maketitle

\section*{Section I: Ordering temperature calculation using the chain random phase approximation (CRPA)}

The dynamical susceptibility for a single Heisenberg chain, $\chi_\mathrm{chain}(\omega,k_\parallel,T)$, calculated from a combination of the Bethe ansatz and field theory techniques, \cite{Bethe1931, Yang1966, Schulz1983, Eggert1994, Schulz1986, Affleck1998, Barzykin2000, Lukyanov1998, Luther1974, TsvelikBook} is
\begin{equation}
\chi_\mathrm{chain}(\omega,k_\parallel, t)
=\Phi(t)\frac{\Gamma\left(\frac{1}{4}-i\frac{\omega-u(k_\parallel-\pi)}{4\pi t}\right)}{\Gamma\left(\frac{3}{4}-i\frac{\omega-u(k_\parallel-\pi)}{4\pi t}\right)}\frac{\Gamma\left(\frac{1}{4}-i\frac{\omega+u(k_\parallel-\pi)}{4\pi t}\right)}{\Gamma\left(\frac{3}{4}-i\frac{\omega+u(k_\parallel-\pi)}{4\pi t}\right)},
\label{eq:chain_chi}
\end{equation}
where $t=k_B T/\jp$, $\Gamma(x)$ is the gamma function, $u=\frac{\pi}{2}\jp b_0$ is the spin velocity, and 
\begin{equation}
\Phi(t)=-\frac{1}{2t}\frac{\sqrt{\ln\left(\frac{\Lambda}{t}\right)}}{(2\pi)^{3/2}}.
\end{equation}
Here, $\Lambda$ is a nonuniversal scale calculated with exact methods by Barzykin to be $\Lambda=24.27\jp/k_B$. \cite{Barzykin2001}

The full three-dimensional dynamical susceptibility within the CRPA is \cite{Scalapino1975, Schulz1996, Essler1997, Bocquet2001}
\begin{equation}
\chi\left(\omega,\bm{k}, T\right)=\frac{\chi_\mathrm{chain}(\omega,k_\parallel, T)}{1-2\tilde{J}_\perp\left(\bm k\right)\chi_\mathrm{chain}(\omega,k_\parallel, T)},
\label{eq:3d_chi}
\end{equation}
where $\tilde{J}_{\perp}(\bm k)$ is the Fourier transform of the inter-chain coupling and $\bm{k}=\left(k_\parallel,k_{\perp 1},k_{\perp 2}\right)$ is the crystal momentum along the respective nearest neighbor bond directions (see Table II of the main text) in units of the crystallographic constants; $a_0=10.277(2)$\,\AA , $b_0=4.6430(9)$\,\AA, and $c_0=11.285(2)$\,\AA\ for the unbent crystal. \cite{Worthy2017} We find, for the frustrated triangular interactions in Fig. 2 of the main text,
\begin{align}
\tilde{J}_{\perp}(\bm k)=&\jone\left[\cos(k_{\perp 1})+\cos(k_{\perp 1}-k_\parallel)\right]\nonumber\\
&+\jtwo\left[\cos(k_{\perp 2})+\cos(k_{\perp 2}-k_\parallel)\right].
\label{eq:Jyz}
\end{align}

One can determine $T_N$ by considering two conditions. Firstly, a zero frequency pole in $\chi\left(0,\bm{k},T\right)$ and secondly, that $\tilde{J}_{\perp}\left(\bm{k}\right)\chi_\mathrm{chain}(0,k_\parallel,T)$ is maximised with respect to $\bm{k}$. That is,
\begin{align}
2\tilde{J}_{\perp}\left(\bm{k}\right)\chi_\mathrm{chain}(0,k_\parallel,T_N)&=1\label{eq:pole_instab1}\\ 
\mathrm{and}\ \ \frac{\partial}{\partial\bm{k}}\left(\tilde{J}_{\perp}\left(\bm{k}\right)\chi_\mathrm{chain}(0,k_\parallel,T_N)\right)&=0.\label{eq:pole_instab2}
\end{align}
These two conditions give both the ordering temperature, $T_N$, and the resulting value of $\bm{k}$, the magnetic ordering wavevector. For a single 1D chain, the maximum in $\chi_\mathrm{chain}\left(0,k_\parallel,T\right)$ occurs at $k=\pi$.

The presence of interchain couplings will shift the ordering wavenumber to an incommensurate value, with the resulting order occurring at $k_\parallel=\pi+k_0$. In \cuacac, there are four values of $k_{\perp 1}$ or $k_{\perp 2}$ (modulo $2\pi$) that satisfy Eq. \ref{eq:pole_instab2}, 
\begin{equation}
 k_{\perp i}=\begin{cases}
 	\pm\frac{|k_\parallel|}{2}, &J_{\perp i}>0 \\
 	\pm\frac{|k_\parallel|}{2} \mp \pi, &J_{\perp i}<0
 	\end{cases}
\end{equation}
In \cuacac, we find $\jone>0$ and $\jtwo<0$, so the possibilities are
\begin{align}
k_{\perp 1}&=\pm\frac{|k_\parallel|}{2},\nonumber\\
k_{\perp 2}&=\pm\frac{|k_\parallel|}{2}\mp \pi.\label{eq:possible_ks}
\end{align}
This gives four possible combinations of $k_{\perp 1}$ and $k_{\perp 2}$. Setting $k_\parallel=\pi+k_0$, we find that, using Eq. \ref{eq:pole_instab2}, all the above possibilities lead to the condition 
\begin{equation}
0=\frac{2\pi T_N}{u|k_0|} + \pi\tanh\left(\frac{|k_0|u}{2k_BT_N}\right) - 2\mathrm{Im}\Psi\left(\frac{1}{4}+i\frac{|k_0|u}{4\pi k_BT_N}\right),
\label{eq:k0}
\end{equation}
which can be solved numerically, yielding
\begin{equation}
\frac{|k_0|u}{4\pi k_BT_N}=\frac{J_\|}{8}\frac{|k_0|}{k_BT_N}\approx0.311.
\label{eq:x}
\end{equation}

Using Eq. \ref{eq:pole_instab1}, we found that the choices in Eq. \ref{eq:possible_ks} yield only two possible solutions for $T_N$. Out of these, we take the highest value of $T_N$, as this is where the instability will occur. This corresponds to $\bm{k}=\left(k_\parallel,|k_\parallel|/2,|k_\parallel|/2-\pi\right)$ or $\bm{k}=\left(k_\parallel,-|k_\parallel|/2,-|k_\parallel|/2+\pi\right)$, which gives
\begin{align}
4\frac{(\jone-\jtwo)}{\jp}&\sin\left(\frac{|k_0|}{2}\right)\chi_\mathrm{chain}(0,k_\parallel,T_N)\nonumber\\
&\approx 0.611\frac{(\jone-\jtwo)}{\jp}
{\sqrt{\ln\left(\frac{\Lambda}{T_N}\right)}}=1
\end{align}
Where we have made a small angle approximation in the second line (this will be strongly vindicated \textit{post hoc}). In units of $\jp$, Eq. \ref{eq:pole_instab1} then becomes
\begin{align} 
T_N
&\approx
\Lambda 
\exp \left[-2.68 \frac{\jp^2}{(\jone+\jtwo)^2} \right] \nonumber\\
&\approx 1.0\times10^{-33} J_\| 
\approx 7.5\times10^{-34}\,\mathrm{K}
\label{eq:TN}
\end{align}
Finally, Eq. \ref{eq:k0} now yields $|k_0|=2.6\times10^{-33}/b_0$, where $b_0$ is the lattice spacing along the chain. Since the value of $k_0$ is so small (vindicating our small angle approximation), the magnetic ordering wavevector along the chain, $k_\parallel$, is approximately $\pi$.

More generally, regardless of the signs of $\jone$ and $\jtwo$, two frustrated interchain couplings will result in ordering at
\begin{equation}
T_N\approx\Lambda\exp\left[-2.68\frac{\jp^2}{\left(|\jone|+|\jtwo|\right)^2}\right].
\label{eq:tn_general}
\end{equation}

When we remove geometric frustration from the lattice (see Fig. 1b of the main text), setting all interchain couplings to be unfrustrated, Eq. \ref{eq:Jyz} becomes
\begin{equation}
\tilde{J}_\perp^\mathrm{cubic}(\bm k)=\jone\cos(k_{\perp 1})+\jtwo\cos(k_{\perp 2})
\label{eq:Jcubic}
\end{equation}
The same CRPA calculation as above then results in 
\begin{equation}
T_N^\mathrm{cubic}\approx 0.56\frac{\left(|\jone|+|\jtwo|\right)}{k_B}\sqrt{\log\left(\frac{\Lambda \jp}{k_BT_N}\right)},
\label{eq:cubic_tn}
\end{equation}

\subsection*{Experimental Predictions}
We use the CRPA to predict a number of experimentally measurable properties of \cuacac. We calculate the 3D susceptibility via Eq. \ref{eq:3d_chi} with two different methods; first with a fit of the CRPA using the Bonner-Fisher expression for the 1D bulk susceptibility and, secondly, using the exact temperature dependent bulk susceptibility of a single antiferromagnetic Heisenberg chain, $\chi_\mathrm{chain}(0,0,T)$,  calculated numerically by Eggert \textit{et al.} \cite{Eggert1994,EggertDat} 

The Bonner-Fisher susceptibility is\cite{Bonner1964, Estes1977, Ami1995}
\begin{equation}
\chi_\mathrm{BF} = \frac{1}{k_BT}\left(\frac{0.25+0.14995x+0.30094x^2}{1+1.9862x+0.68854x^2+6.0626x^3}\right)
\end{equation}
where $x=|\jp|/(2k_BT)$. We fit the expression 
\begin{equation}
\chi_\mathrm{expt}(T)=Ng^2\mu_B^2\left(\frac{\chi_\mathrm{BF}(T)}{1-2J_\perp^\ast\chi_\mathrm{BF}(T)}\right)\label{eq:chiBF}
\end{equation}
where the Land\'e g-factor $g$ and the interchain coupling $J_\perp^\ast$, are free parameters and $\jp$ is fixed to our BS-DFT result, 0.75\,K. Our fit to the experimental data between 2\,K and 10\,K from Moreno \textit{et al.} \cite{Moreno2013} results in 
\begin{align*}
g&=2.1,\\
J_\perp^\ast&=0.28\ \mathrm{K}.
\end{align*}
The result for $g$ is a typical value for other cuprates. \cite{Ami1995} The value $J_\perp^\ast$ corresponds to $\bm{k}=(\bm 0)$ in Eq. \ref{eq:Jyz}, so our result predicts $\tilde{J}_{\perp}(\bm 0)=2\left(\jone+\jtwo\right)=0.28$\,K. This is comparable with our BS-DFT results. 

We then calculated the low-temperature prediction using these parameters as 
\begin{equation}
\chi_\mathrm{pred}(T)=Ng^2\mu_B^2\left(\frac{\chi_\mathrm{chain}(0,0,T)}{1-2J_\perp^\ast\chi_\mathrm{chain}(0,0,T)}\right)\label{eq:chiEgg}
\end{equation}
with $\chi_\mathrm{chain}(0,0,T)$ from Eggert \textit{et al.} \cite{Eggert1994,EggertDat} 

The magnetic susceptibility is related to the dynamical structure factor, which is measured in inelastic neutron-scattering experiments, by \cite{Lovesey}
\begin{equation}
S(\omega,\bm{k},T)=-\frac{1}{1-\exp\left(-\omega/T\right)}\mathrm{Im}\chi(\omega,\bm{k},T).
\label{eq:structure_fac}
\end{equation}

\section*{Section II: Generation of Linearized Crystals}

\begin{figure}
\centering
\includegraphics[width=8cm]{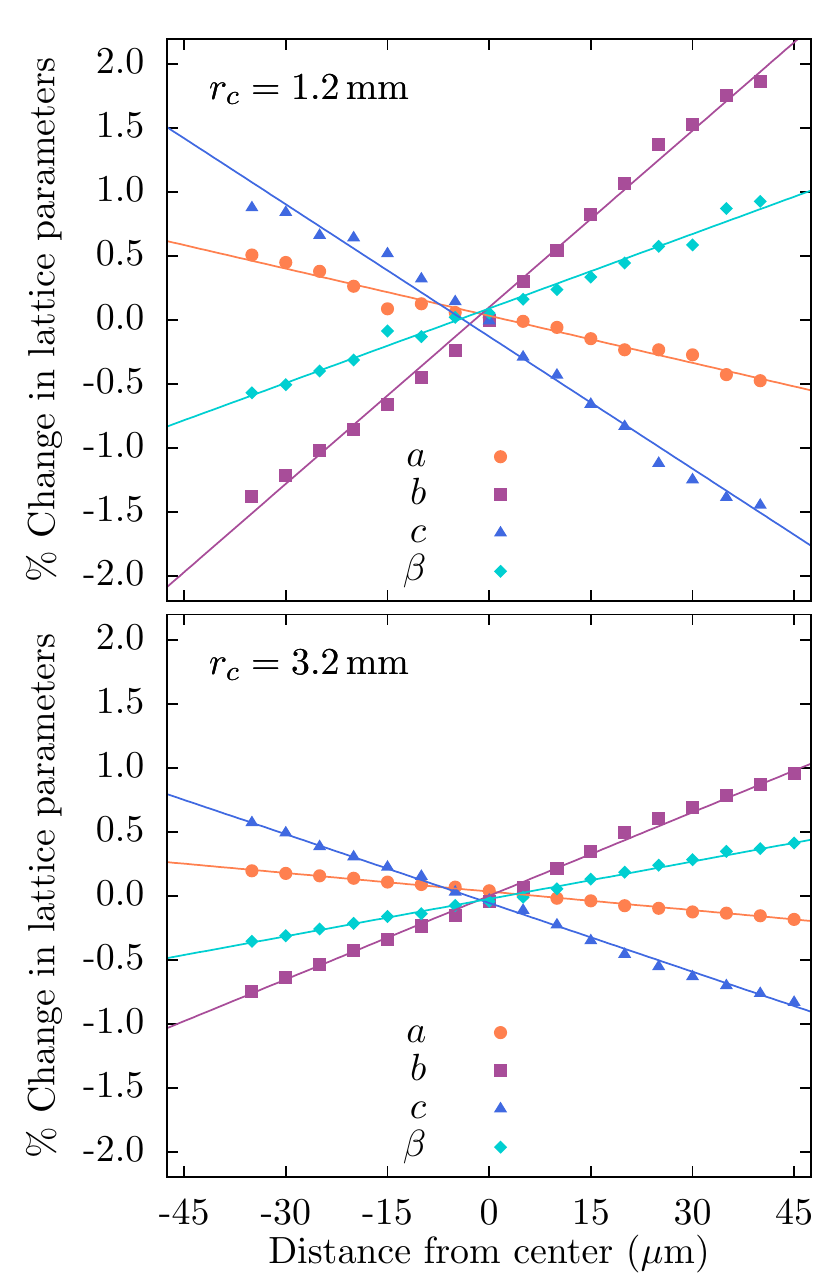}
\caption{Linear regressions (Eqs. \ref{eq:reg_1} and \ref{eq:reg_2}) of the change in lattice parameters across both crystals. The data points are from Worthy \textit{et al.} \cite{Worthy2017} The change in lattice parameters is relative to the unbent crystal. The center is defined as the point in the each bent crystal where $b$ is closest to the unbent $b_0$. The data are labelled by the radius of curvature, $r_c$.}
\label{fig:latt_prams}
\end{figure}

Figure \ref{fig:latt_prams} shows the lattice parameters of the bent crystals, from Worthy \textit{et al.}, \cite{Worthy2017} along with our linear regressions. For radius of curvature $r_c=1.2$\,mm, we found
\begin{align}
\frac{a^\prime}{a_0} &= -0.00061d + 0.0052\nonumber\\
\frac{b^\prime}{b_0} &= 0.00230d - 0.0174\nonumber\\
\frac{c^\prime}{c_0} &= -0.00172d + 0.0125\nonumber\\
\frac{\beta^\prime}{\beta_0} &= -0.00097d - 0.0069,
\label{eq:reg_1}
\end{align}
where $d$ is the distance from the center of the crystal (defined to be where the data for $b^\prime$ is closest to $b_0$) in $\mu$m. For $r_c=3.2$\,mm,
\begin{align}
\frac{a^\prime}{a_0} &= -0.00024d + 0.0023\nonumber\\
\frac{b^\prime}{b_0} &= 0.00109d - 0.0087\nonumber\\
\frac{c^\prime}{c_0} &= -0.00089d + 0.0066\nonumber\\
\frac{\beta^\prime}{\beta_0} &= -0.00049d - 0.0041.
\label{eq:reg_2}
\end{align}
Using these regressions, we produced a linear set of lattice parameters for points across the bent crystals. In order to minimise noise in our predictions, we then produced a set of nearest neighbor dimers to use as input for our BS-DFT calculations. These structures are attached as `bent\_dimers.zip'. To correctly simulate the change of the dimer coordinates across the bends, we kept the intramolecular bond lengths constant by producing new Wyckoff coordinates, $(x^\prime, y^\prime, z^\prime)$, for each dimer using the Wyckoff coordinates from the unbent crystal structure, $(x^0, y^0, z^0)$, measured by Worthy \textit{et al.};\cite{Worthy2017}
\begin{equation}
\begin{pmatrix}
x^\prime_1 & x^\prime_2 &  \ldots \\
y^\prime_1 & y^\prime_2 &  \ldots \\
z^\prime_1 & z^\prime_2 &  \ldots 
\end{pmatrix}
=
\begin{pmatrix}
\frac{1}{a^\prime} & 0 & -\frac{1}{a^\prime}\cot(\beta^\prime) \\
0 & \frac{1}{b^\prime} & 0\\
0 & 0 & \frac{1}{c^\prime}\csc(\beta^\prime)
\end{pmatrix}
\begin{pmatrix}
a_0 & 0 & c_0\cos(\beta_0)\\
0 & b_0 & 0\\
0 & 0 & c_0\sin(\beta_0)
\end{pmatrix}
\begin{pmatrix}
x^0_1 & x^0_2 &  \ldots \\
y^0_1 & y^0_2 &  \ldots \\
z^0_1 & z^0_2 &  \ldots 
\end{pmatrix},
\label{eq:matrix_transform}
\end{equation}
where $a^\prime$, $b^\prime$, $c^\prime$, and $\beta^\prime$ are the new, linearized crystallographic paramaeters. The Wyckoff coordinates include the coordinates for one copper atom and one acetylacetonate (acac) unit. After transforming the Wyckoff coordinates as above, we created the other coordinates (one more copper atom and three more acac units) using the symmetry transformations of the crystal space group, $P2_1/c$. For all dimers, the first molecule is the Wyckoff coordinates, $(x,y,z)+(0,0,0)$, along with the second acac unit given by $(-x,-y,-z)+(1,1,1)$. Coordinates for the second molecule (third and fourth acac units, along with one copper atom) are given by their relevant symmetry transformations,
\begin{align*}
\mathrm{Along\ chain}:\ &(x,y,z)+(0,1,0)\ \mathrm{and}\\
&(-x,-y,-z)+(1,2,1)\\
\mathrm{Along }\perp 1:\  &(-x+0.5,y+0.5,-z+0.5)+(0,0,0)\ \mathrm{and}\\
& (x+0.5,-y+0.5,z+0.5)+(-1,1,-1)\\
\mathrm{Along }\perp 2:\  &(-x+0.5,y+0.5,-z+0.5)+(1,0,0)\ \mathrm{and}\\
& (x+0.5,-y+0.5,z+0.5)+(0,1,-1).
\end{align*}

We chose this method to make the bond distances in our linearized crystals the same as the unbent structure, while the distances and angles between molecules changed; this is how the coordinates change in the original crystollographic data. \cite{Worthy2017}

\section*{Section III: BS-DFT Benchmarking}

Table \ref{tab:basis_benchmark} shows BS-DFT results for all three nearest neighbour couplings with different basis sets. The magnitudes of the interchain couplings are all very similar and the range of $J_\parallel$ values is 0.15\,K, which is 20\% of our reported value $J_\parallel=0.75$\,K. 

As shown in Table \ref{tab:func_benchmark}, changing the functional between a pure functional, PBE, a hybrid, B3LYP, and a hybrid meta-GGA, uTPSSh, results in a larger range of couplings, which is expected. However, these three functionals still result in a large change in couplings across a bend (Table \ref{tab:bent_benchmark}).

Importantly, all results show that the magnetic exchange model of \cuacac\ is highly one-dimensional. There are two cases, in Tables \ref{tab:basis_benchmark} and \ref{tab:func_benchmark}, where $J_{\perp 1}$ changes sign. This does not change our result, as $T_N$ depends only on the magnitudes of the interchain couplings (see Eq. \ref{eq:tn_general}).

\begin{table}
\caption{Heisenberg exchange ($J_{ij}$) parameters for the unbent structure of \cuacac\, determined with BS-DFT using different basis sets. All calculations were performed in Gaussian09 \cite{g09} with the uB3LYP functional \cite{B3LYPa,B3LYPb} and an SCF convergence criterion of $10^{-10}$\,a.u. }\label{tab:basis_benchmark}
\begin{center}
\begin{tabular}{c|ccc}
Basis Set  & $J_\parallel$\,(K) & $J_{\perp 1}$\,(K)  & $J_{\perp 2}$\,(K)\\ 
\hline 
6-31+G*\cite{Pople1, Pople2, Pople3, Pople4, Pople_3rdRow} & 0.73 & -0.05 & -0.08\\ 
6-31++G**\cite{Pople1, Pople2, Pople3, Pople4, Pople_3rdRow} & 0.75 &  0.05 & -0.07  \\ 
6-311+G\cite{6311Ga, 6311Gb, 6311Gc, 6311Gd} & 0.67 &  0.04 & -0.08 \\ 
6-311+G(3df,p)\cite{6311Ga, 6311Gb, 6311Gc, 6311Gd} & 0.71 &  0.03 & -0.09 \\ 
TZVP\cite{TZVPa, TZVPb} & 0.76 & 0.03 & -0.08 \\
6-31+G*\cite{Pople1, Pople2, Pople3, Pople4}  and LANL2DZ\cite{Dunning1977, Hay1985a, Wadt1985, Hay1985b} (Cu) & 0.75 &  0.04 & -0.1 \\
TZVP\cite{TZVPa, TZVPb} and LANL2DZ\cite{Dunning1977, Hay1985a, Wadt1985, Hay1985b} (Cu) & 0.82 & 0.03 & -0.09 \\
aug-cc-pVTZ\cite{Dunning1, Dunning2, Dunning3} & 0.70 & 0.04 & -0.06 \\
\end{tabular}
\end{center}
\end{table}

\begin{table}
\caption{Heisenberg exchange ($J_{ij}$) parameters for the unbent structure of \cuacac\, determined with BS-DFT using different exchange-correlation functionals. All calculations were performed in Gaussian09 \cite{g09} using the LANL2DZ\cite{Dunning1977, Hay1985a, Wadt1985, Hay1985b} (for Cu) and 6-31+G*\cite{Pople1, Pople2, Pople3, Pople4} basis sets with an SCF convergence criterion of $10^{-10}$\,a.u. }\label{tab:func_benchmark}
\begin{center}
\begin{tabular}{c|ccc}
Functional  & $J_\parallel$\,(K) & $J_{\perp 1}$\,(K)  & $J_{\perp 2}$\,(K)\\ 
\hline 
uPBEPBE\cite{PBEa, PBEb} & 1.4 & 0.09 & -0.21\\ 
uB3LYP\cite{B3LYPa,B3LYPb} & 0.75 &  0.04 & -0.10  \\ 
uTPSSh\cite{TPSSha, TPSShb} & 0.60 &  -0.04 & -0.14
\end{tabular}
\end{center}
\end{table}

\begin{table}
\caption{Intrachain Heisenberg exchange ($J_\parallel$) for the inside and outside of the bent crystal with $r_c=1.2$\,mm using different functionals and basis sets. All calculations were performed in Gaussian09 \cite{g09} with an SCF convergence criterion of $10^{-10}$\,a.u. }\label{tab:bent_benchmark}
\begin{center}
\begin{tabular}{c|c|ccc}
Functional  & Basis Set & In. $J_\parallel$\,(K) & Out. $J_\parallel$\,(K)  & \% Change\\ 
\hline 
uPBEPBE\cite{PBEa, PBEb} & 6-311+G\cite{6311Ga, 6311Gb, 6311Gc, 6311Gd} & 1.02 & 0.52 & 49\\
 & 6-31+G*\cite{Pople1, Pople2, Pople3, Pople4}, LANL2DZ\cite{Dunning1977, Hay1985a, Wadt1985, Hay1985b} (Cu) & 1.25 & 0.67 & 46 \\
 & TZVP\cite{TZVPa, TZVPb}, LANL2DZ\cite{Dunning1977, Hay1985a, Wadt1985, Hay1985b} (Cu) & 1.36 & 0.78 & 44\\ 
 \hline
uB3LYP\cite{B3LYPa,B3LYPb} & 6-311+G\cite{6311Ga, 6311Gb, 6311Gc, 6311Gd} & 0.64 & 0.43 & 33\\
 & 6-31+G\cite{Pople1, Pople2, Pople3, Pople4}, LANL2DZ\cite{Dunning1977, Hay1985a, Wadt1985, Hay1985b} (Cu)  & 0.69 &  0.46 & 33  \\ 
  & TZVP\cite{TZVPa, TZVPb}, LANL2DZ\cite{Dunning1977, Hay1985a, Wadt1985, Hay1985b} (Cu)  & 0.75 &  0.49 & 34  \\ 
   \hline
uTPSSh\cite{TPSSha, TPSShb} & 6-311+G\cite{6311Ga, 6311Gb, 6311Gc, 6311Gd} & 0.39 & 0.16 & 57\\
 & 6-31+G*\cite{Pople1, Pople2, Pople3, Pople4}, LANL2DZ\cite{Dunning1977, Hay1985a, Wadt1985, Hay1985b} (Cu)  & 0.53 &  0.25 & 52  \\ 
  & TZVP\cite{TZVPa, TZVPb}, LANL2DZ\cite{Dunning1977, Hay1985a, Wadt1985, Hay1985b} (Cu)  & 0.58 &  0.31 & 47  \\ 
\end{tabular}
\end{center}
\end{table}

\pagebreak
\bibliography{cu_acac}